\documentclass[10pt, conference, letterpaper]{IEEEtran}
\usepackage{graphicx}
\usepackage{amsfonts,amsmath}
\usepackage{color,colordvi}
\usepackage{textcomp}
\usepackage{epsfig}
\usepackage{url}
\usepackage{mdwlist}
\usepackage{subfigure}
\usepackage{verbatim}
\usepackage{multirow,multicol}
\usepackage{setspace}

\usepackage{float}
\usepackage{algorithm}
\usepackage{algorithmic}
\usepackage{enumerate}
\usepackage{fancyhdr}
\usepackage[normalem]{ulem}

\newcommand{\para}[1]{\vspace{5pt}\noindent{\bf #1.}}

\newlength{\saveparindent}
\newlength{\saveparskip}
\newtheorem{HNRATheo}{Theorem}
\newtheorem{HNRALemma}{Lemma}

\begin{document}

%\conferenceinfo{ICN'14,} {February 18--20, 2013, San Antonio, Texas, USA.}
%\CopyrightYear{2013}
%\crdata{978-1-4503-1890-7/13/02}
%\clubpenalty=10000
%\widowpenalty = 10000

\title{ELDA: Towards Efficient and Lightweight Detection of Cache Pollution Attacks in NDN}
%
%\author{
%Zhiwei Xu$^{\dag}$, Bo Chen$^{\ddag}$, Yujun Zhang$^{\dag}$, Zhongcheng Li$^{\dag}$, Xin Wang$^{\S}$\\
%$^{\dag}$Institute of Computing Technology, Chinese Academy of Sciences, Beijing, P.R.China\\
%$^{\ddag}$Dept. of Computer Science, New Jersey Institute of Technology, Newark, USA\\
%$^{\S}$Dept. of Electrical and Computer Engineer, Stony Brook University, New York, USA\\
%Email: \{xuzhiwei2001,zhmj,zcli\}@ict.ac.cn, bc47@njit.edu, xwang@ece.sunysb.edu
%}
\author{
Zhiwei Xu$^{\dag}$$^{\ddag}$, Bo Chen$^{\S}$, Ninghan Wang$^{\dag}$$^{\ddag}$, Yujun Zhang$^{\ddag}$, Zhongcheng Li$^{\ddag}$\\
\small $^{\dag}$University of Chinese Academy of Sciences, Beijing, China\\
\small $^{\ddag}$Institute of Computing Technology, Chinese Academy of Sciences, Beijing, China\\
\small $^{\S}$Department of Computer Science, Stony Brook University, New York, USA\\
\small Email: \{xuzhiwei2001,wangninghan,zhmj,zcli\}@ict.ac.cn, bochen1@cs.stonybrook.edu
}
\maketitle
\thispagestyle{empty}
%\pagenumbering{arabic}

\begin{abstract}
As a promising architectural design for future Internet, named data networking (NDN) relies on in-network caching to efficiently deliver name-based content. However, the in-network caching is vulnerable to cache pollution attacks (CPA), which can reduce cache hits by violating cache locality and significantly degrade the overall performance of NDN.

To defend against CPA attacks, the most effective way is to first detect the attacks and then throttle them. Since the CPA attack itself has already imposed a huge burden on victims, to avoid exhausting the remaining resources on the victims for detection purpose, we expect a lightweight detection solution. We thus propose ELDA, an Efficient and Lightweight Detection scheme against cache pollution Attacks, in which we design a Lightweight Flajolet-Martin (LFM) sketch to monitor the interest traffic. Our analysis and simulations demonstrate that, by consuming a few computation and memory resources, ELDA can effectively and efficiently detect CPA attacks.
\end{abstract}

%\keywords{ Hierarchical name; Route aggregation; Coding polynomial; Compact representation; Named Date Networking (NDN)}

\section{Introduction}
\label{sec:introduction}

In the past several decades, Internet has evolved from a simple communication network for point-to-point conversations to a sophisticated global system that supports communication needs of all types of services, including packet delivery for information-intensive business, e-commerce and contents \cite{gantz2012digital,roberts2009clean}. Existing Internet architecture relies on IP address to guide packet transmissions. However, originally designed to suit the need of connection driven end-to-end communications, this transmission infrastructure is inefficient in supporting content oriented applications. As a promising architectural design for the future Internet, NDN \cite{jacobson2009networking, Zhang2010data} was proposed to bridge the gap between the current Internet architecture and the information centric new applications. NDN leverages in-network caching as well as multipath communication to implement the name-based content (data) retrieval effectively and efficiently. 
 
Unfortunately, in-network caches may become victims of malicious behaviors. A malicious user can violate the content locality of the in-network caches by performing a cache pollution attack (CPA)~\cite{gupta1990improving,gao2006internet}. In this way, it can cause a large number of cache misses and significantly degrade quality of services for regular users. There are primarily two types of CPA attacks: locality-disruption attack (LDA) and false-locality attack (FLA)~\cite{gao2006internet}. An LDA attack continuously requests distinct unpopular content, by which it keeps pushing the unpopular content into the cache of the routers along the transmission path, and thus squeezing out the popular content. To perform an FLA attack, the malicious user repeatedly requests the same set of unpopular content, by which it is able to keep this set of unpopular content in the cache, and thus reduce cache hits of content requests from legitimate users. 

In NDN, it is difficult for a malicious user to perform the FLA attack in a conventional way. 
When a piece of content is retrieved, the routers on the transmission path will cache a copy of this content, which will be used to satisfy subsequent requests for this content~\cite{Zhang2010data}. Therefore, the repeated malicious requests for the same content will be satisfied by the copy cached in the neighboring routers, and will not be forwarded to other routers. Thus, repeatedly sending malicious requests for the same content can only increase this content's cache hits in the neighboring routers, \emph{i.e.,} a conventional FLA attack can only affect the neighboring routers of the malicious users, rather than the overall NDN network. However, we successfully construct an FLA attack which can affect the whole NDN network. In our construction, we smartly throttle the newly arriving content to prevent the unpopular content from being squeezed out. %Specifically, the attacker first requests distinct unpopular content, such that the unpopular content will fill up the in-network cache. It then sends a large number of requests for nonexistent content in order to block newly arriving content (i.e., an interest flooding attack \cite{gasti2013and}). Consequently, the unpopular content will stay in the cache.

A CPA attack will significantly degrade the performance of an NDN network. To defend against this attack, several countermeasures are proposed in the literature. The existing proactive countermeasure \cite{xie2012enhancing} mitigates CPA attacks with the cost of disturbing regular cache usage. To throttle the attack traffic accurately, multiple approaches are proposed to detect CPA attacks \cite{conti2013lightweight, karami2015anfis}. However, both  their time and space complexity are too high. When a CPA attack happens, the attack itself has already imposed a huge burden on the victim. Thus, a detection solution should be kept as lightweight as possible to avoid consuming too many additional resources and overwhelming the victim. Meanwhile, the detection scheme should be efficient enough to capture the symptom of attack traffic at wire speed, when maintaining high detection accuracy.

%We observe that to perform either an LDA attack or an FLA attack, the attacker needs to send a large number of interests for distinct content in a short period, which will cause a significant difference between the attack traffic and the regular traffic in NDN. We rely on this symptom to detect the CPA attacks.
To design an efficient and lightweight detection mechanism, we face multiple challenges:
\begin{enumerate}
\item The attacker may mimic legitimate requests, thus the malicious requests are indistinguishable from the legitimate ones. To identify CPA attacks, we need to analyze these attacks to choose suitable symptoms for detection purpose.
\item The attackers usually send a huge number of malicious interests in a short period to perform CPA attacks. How to accurately and timely capture the symptom of the attack traffic is a challenge.
\item The CPA attack itself has already consumed a large number of computation and memory resources of the victim. Utilizing minimal computation and memory resources from the victim to detect CPA attacks is not straightforward.
\end{enumerate}

To handle the aforementioned challenges, we investigate the impact of CPA attacks in NDN and propose a lightweight detection solution. Our solution takes advantage of the difference between the attack traffic and the regular traffic. We demonstrate that our approach provides a better performance at a lower cost, and is able to detect both the LDA and the FLA attack accurately. 

\para{Contributions} Our contributions are summarized as follows.

\begin{enumerate}
%\item We successfully construct a novel FLA attack by throttling the newly arriving content to prevent the unpopular content from being squeezed out.
\item We analyze both the LDA and the FLA attack, and discover a significant symptom in both attacks, by which we design a across-the-board scheme to detect both of them.
\item We design a lightweight Flajolet-Martin sketch, namely, LFM sketch, which can capture the symptom of attack traffic effectively and efficiently. In addition, we construct ELDA, an efficient CPA detection solution, which uses the LFM sketch to monitor interest traffic with high performance and low resource consumption.
\item We analyze the security of ELDA, and evaluate its performance in ndnSIM~\cite{afanasyev2012ndnsim}. 
\end{enumerate}

In Section \ref{sec:background}, we provide thorough background knowledge on NDN and introduce the primitives we rely on in our design. We then present our threat model in Section \ref{sec:threatmodels}. We design an efficient scheme to detect CPA attacks in Section \ref{sec:scheme}, and analyze its security  in Section \ref{sec:analysis}. We study CPA attacks in NDN and evaluate the effectiveness of our design in Section \ref{sec:evaluations}. We outline the related work in section \ref{sec:related} and conclude in Section \ref{sec:conclusion}.

\section{Background}
\label{sec:background}
\subsection{Named Data Networking}

Unlike IP-based architecture, NDN uses hierarchically structured names rather than IP addresses to guide packet routing and forwarding. 
%NDN leverages in-network as well as multipath communication to implement this name-based content (data) retrieval effectively and efficiently and further support various functionality, including content distribution, multicast, mobility, and delay-tolerant networking \cite{Zhang2010data}. 
%In the following, we briefly introduce NDN with the basic background.

%NDN uses two types of packets: interest packet (data request packet) and content packet (data response packet).The interest packet and the corresponding content packet are identified by the same hierarchical name.

%\textbf{Router architecture:}An NDN router includes the following three components: (1) Pending Interest Table (PIT) is used to record interest forwarding information, keeping track of unsatisfied interests and corresponding ingress interfaces; (2) Forwarding Information Base (FIB) maintains an interest forwarding table of name prefixes and corresponding egress interfaces; (3) Content Store (CS) is used to cache the received content.

An NDN user retrieves name-based content by three steps: (1) To request name-based content (data), a consumer (user) first sends an interest (data request packet) identified by the name of the requested content. (2) After having received this interest packet, a router records this interest and its ingress interface into its PIT (Pending Interest Table). Note that the duplicate interests are aggregated to one entry by appending the ingress interface list. If the requested content has been cached in its CS (Content Store), it will satisfy this interest with the cached copy. Otherwise, it will search its FIB (Forwarding Information Base) to longest match this interest's name and find the egress interfaces, and forward this interest from these interfaces to the next routers. If this interest can not be satisfied by the routers along its transmission path, it will reach the corresponding content provider (content producer). (3) When the content packet is sent back, the router finds the corresponding PIT entry and forwards the content packet to all the ingress interfaces listed in the PIT entry. In addition, the router removes the corresponding PIT entry, and caches the content packet in its CS.

\subsection{Primitives}
\subsubsection{Flajolet-Martin (FM) Sketch}
\label{sec:FM_sketch}
\begin{figure}[H]
\setlength{\abovedisplayskip}{3pt}
\setlength{\belowdisplayskip}{3pt}
\label{fig:FM_sketch}
%\vspace{-10pt}
\centering
\includegraphics[width=0.45\textwidth]{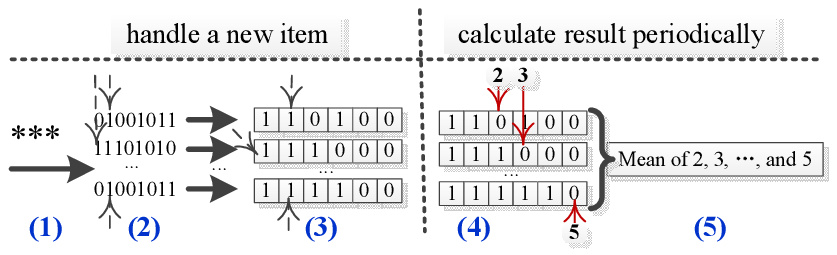}
\vspace{-5pt}
\caption{The process of the FM sketch}
\vspace{-10pt}
\label{fig:The_process_of_the_FM_sketch}
\end{figure}

Flajolet et al. \cite{flajolet1985probabilistic} originally propose FM sketch, which is a bitmap-based counting algorithm to estimate the number of distinct items (\emph{i.e.}, cardinality) among numerous items at a cost of small memory consumption. The FM sketch first selects a set of hash functions, and constructs for each hash function a corresponding bitmap. Note that the size of the bitmap is equal to the size of the hash value generated by the hash function. The FM sketch then works as follows (see the example in Fig.\ref{fig:The_process_of_the_FM_sketch}): 
(1) It applies the set of hash functions over each scanned item, generating a set of hash values; (2) It searches the leftmost ``$1$'' bit in each hash value; (3) It aggregates the search result into a bitmap by setting the bit on the same position of the bitmap to be ``$1$''. Note that a bitmap is used to keep track of the position of the leftmost ``1'' bit in the hash values generated by the corresponding hash function; (4) It searches the leftmost ``$0$'' bit in each bitmap; (5) It calculates the arithmetic mean of the indices of the leftmost ``$0$'' bits to estimate the cardinalities of the scanned items. To handle a new item, the FM sketch performs step (1) to (3), and to obtain the estimation result, the FM sketch performs step (4) to (5). The latest version of FM sketch, the hyperloglog FM sketch \cite{flajolet2008hyperloglog}, is the most effective FM sketch, which uses harmonic mean rather than the arithmetic mean.

\subsubsection{Monte Carlo hypothesis test}
Monte Carlo hypothesis test \cite{metropolis1949monte} constructs several repeated tests and estimates the underlying distribution according to the test results. To provide sufficient samples for all the repeated tests, Monte Carlo hypothesis test re-uses the latest sample by randomly re-sampling with replacement from this sample (\emph{i.e.}, Bootstrapping). Therefore, Monte Carlo hypothesis test needs less samples to estimate the underlying distribution compared to a conventional hypothesis test. We perform a Monte Carlo hypothesis test by four steps: (1) We acquire a sample of size $n$ from the interested data; (2) We randomly select an observation on the sample as a resample, record its value, and put it back so that it has a chance to be drawn again. In this way, we generate several resamples of size $n$ from the original sample; (3) We compute an estimation of the interested data from each resample, and display all the estimations computed from the resamples in a histogram, which represents the sampling distribution of the interested data; (4) we obtain an appropriate estimation of the interested data according to the aforementioned distribution and the selected level of significance.
\section{Threat Model}
\label{sec:threatmodels}
The CPA attacks usually target to violate the content locality in NDN network caches. When a CPA attack is successfully performed, the legitimate users will not be able to retrieve the expected content from the network caches. Instead, they may need to turn to the distant content provider, which will bring higher delay and reduce network throughput. CPA attacks can be characterized into two types, a locality-disruption attack (LDA) and a false-locality attack (FLA). An LDA attack continuously requests distinct unpopular content, ruining the cache content locality. In contrast, an FLA attack repeatedly requests the same set of unpopular content, creating a false content locality at caches.

Since an NDN router will discard repeated interests due to the interest aggregation mechanism in its PIT, a conventional FLA attack is difficult to be performed, which requires the attacker to repeatedly solicit the same set of content.
However, we successfully construct a new FLA attack by two steps: %(see Fig. \ref{fig:Attack Model for FLA}):
(1) The attacker requests distinct unpopular content to deplete the content stores of the routers along the critical transmission path (\emph{e.g.,} gateway routers). (2) The adversary sends a large number of interests with forged names to overwhelm those routers' PIT tables (\emph{i.e.}, an interest flooding attack~\cite{gasti2013and}). In this way, newly arriving interests will be dropped due to the insufficiency of PIT space, and thus no new content will arrive. Consequently, the unpopular content will stay in those routers' content stores.
%\begin{figure}[H]
%\label{fig:Attack Model}
%\vspace{-15pt}
%\centering
%\includegraphics[width=0.45\textwidth]{./figs/f1}
%\vspace{-10pt}
%\caption{Attack Model for FLA}
%\vspace{-10pt}
%\label{fig:Attack Model for FLA}
%\end{figure}

To enhance the effect of both the LDA and the FLA attack, an attacker usually needs to require the compromised consumers to send an extremely large number of malicious interests (\emph{i.e.,} interests requesting unpopular content or nonexistent content) during a short period. In addition, for both the LDA and the FLA attack, when sending out malicious interests, the compromised consumers will use the same name prefix. In this way, the malicious interests will be forwarded towards the content provider corresponding to the targeted name prefix~\cite{Zhang2010data}, and will converge on the routers along the critical transmission path in a short period, which will lead to a situation that the responded content will deplete those routers' content stores. In summary, to perform an effective CPA attack, the attacker always needs to send a large number of malicious interests in a short period, requesting distinct content whose names possess the same prefix. Therefore, when a victim monitors the interests which possess the same name prefix used during a CPA attack, it can find that the number of the distinct content requested by the these interests dramatically increases. This is a significant symptom which can be used to identify CPA attacks.

\section{An Efficient and Lightweight Detection of CPA Attacks in NDN (ELDA)}
\label{sec:scheme}

In this section, we propose ELDA, an efficient and lightweight scheme for detecting CPA attacks. Specifically, in Section \ref{Strawman Solution}, we provide a strawman solution, in which we directly rely on the FM sketch (see Section~\ref{sec:FM_sketch}). 
This simple solution, however, is not efficient enough due to the large number of redundant hash operations performed in the FM sketch. In Section \ref{ELDA}, we propose a novel lightweight FM (LFM) sketch, and design ELDA based on this LFM sketch.

\subsection{A Strawman Solution and Its Limitations}
\label{Strawman Solution}
To throttle the CPA attacks accurately in NDN, we need to find out the name prefixes used by the attacker for sending malicious interests. When a router monitors the interests possessing a common name prefix, it can observe that the number of distinct content requested by these interests would increase dramatically when the attacker performs a CPA attack on it (Section~\ref{sec:threatmodels}). In general, we can rely on this symptom to detect CPA attacks in two phases, a traffic monitoring phase and an attack identification phase. In the traffic monitoring phase, the router monitors interests possessing a common name prefix, such that it can find the burst of distinct content requested by these interests. The FM sketch (Section~\ref{sec:FM_sketch}) is designed for estimating the number of distinct items among numerous scanned items. Thus, in this phase, the router can simply rely on the FM sketch to monitor the interests which possess a common name prefix, such that it can find the burst of the distinct content requested by these interests.

In the attack identification phase, the router periodically obtains the interest traffic monitoring result (\emph{i.e.,} the estimation result of the FM sketch), and compares this result with an empirical threshold.
If the result exceeds the threshold, the router reports a potential CPA attack immediately. 

\para{Limitations} To guarantee an accurate traffic monitoring result, the strawman solution needs to use a large set of hash functions~\cite{flajolet1985probabilistic}. Even worse, all the hash functions need to be called when a new interest is received, and then all the corresponding bitmaps need to be updated. This requires a large amount of computation and memory resources.

\subsection{ELDA}
\label{ELDA}
\subsubsection{A Lightweight FM Sketch (LFM sketch)}
\label{subsubsec:FM sketch}

To minimize the resource consumption during the monitoring phase, we design a novel lightweight Flajolet-Martin sketch (LFM sketch), which can efficiently estimate the cardinality of the requested content. 

\begin{figure}[H]
\label{fig:lightweight_FM_sketch}
  \vspace{-10pt}
  \centering
    \includegraphics[width=0.45\textwidth]{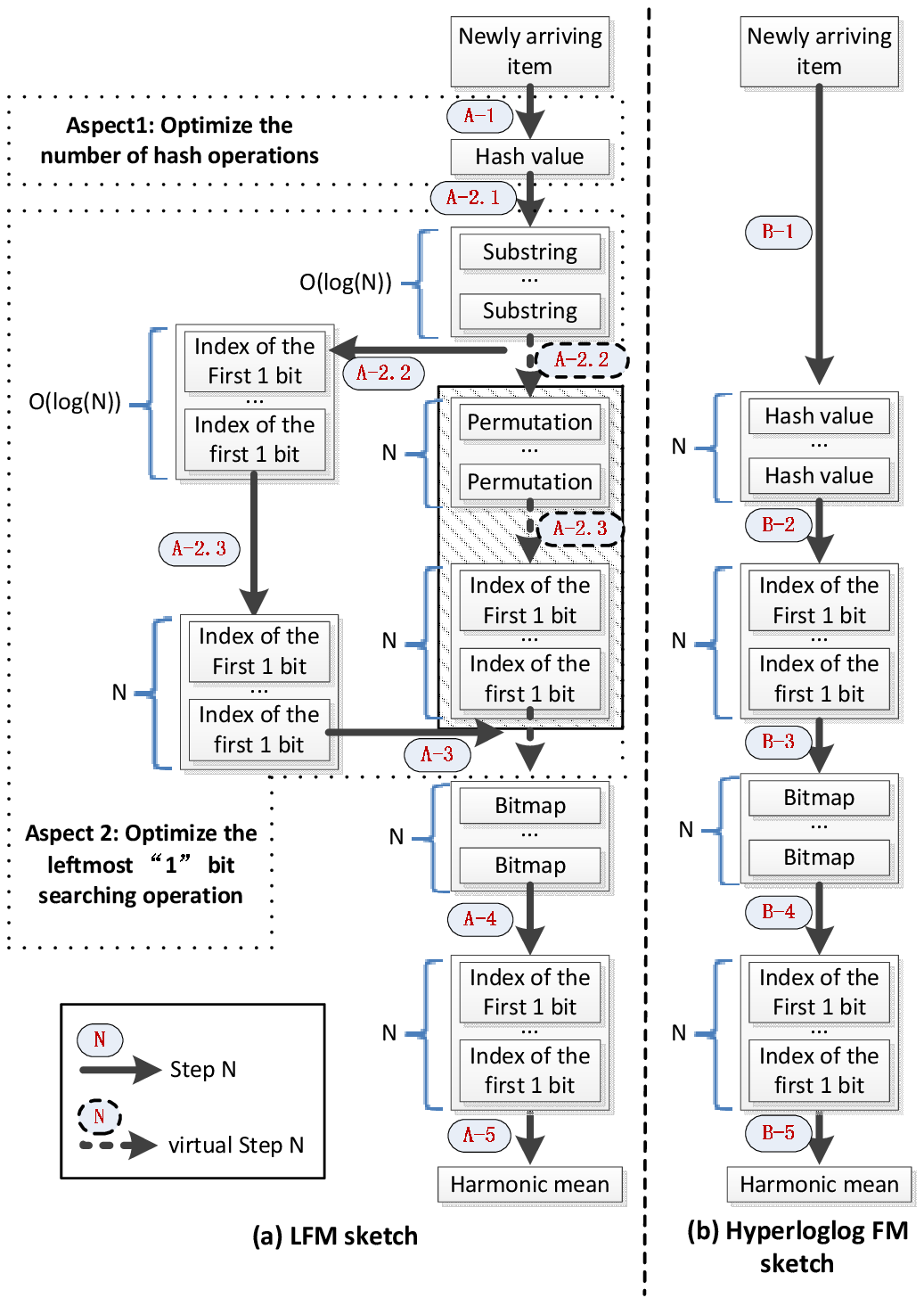}
  \vspace{-10pt}
  \caption{The process of the LFM sketch}
  \vspace{-10pt}
  \label{The process of the lightweight FM sketch}
\end{figure}

Since the conventional FM sketch (\emph{e.g.,} the hyperloglog FM sketch) only uses a small fraction of information in each generated hash value (\emph{i.e.,} the leftmost ``1’’ bit), it requires to generate a large number of hash values. \textbf{Our novel LFM sketch tries to fully utilize all the bits of a hash value, such that we can eliminate the unnecessary overhead resulted from the hash operations.} 
Our LFM sketch outperforms the existing FM sketch in two aspects. Firstly, we perform less number of hash operations during the monitoring phase.
Our key idea is to use only one hash function instead of a set of hash functions.
For each scanned item, we first use a hash function to generate a hash value, and then re-arrange the bits in this hash value, generating a large number of permutations, which can be used to replace the hash values generated by multiple hash functions in the existing FM sketch. As it is shown in Section~\ref{sec:analysis}, by using one hash function in the aforementioned way, we achieve the similar estimation accuracy as using multiple hash functions. Secondly, we further optimize our LFM sketch by skipping the permutation generation process. We present our LFM sketch in Fig. \ref{The process of the lightweight FM sketch} (For convenience, we also include the steps of the hyperloglog FM sketch as a comparison), with a detailed explanation in the following.

For each incoming item, we first generate a hash value (step A-1). Compared to the hyperloglog FM sketch algorithm (step B-1), our LFM sketch significantly reduces the number of hash operations. To guarantee estimation accuracy, we need an equal number of random values comparable to the hash values in the hyperloglog FM sketch. Thus, we split the hash value into a group of substrings (step A-2.1), and re-arrange these substrings to generate their permutations (virtual step A-2.2), which will be used as the hash values in the hyperloglog FM sketch. We further update a bitmap (step A-3) by setting its $j$-th bit as 1, where $j$ is the index of the leftmost ``1'' bit of the corresponding permutation. Periodically, we obtain the index of the leftmost ``0'' bit in each bitmap, and compute the harmonic mean of these indices, estimating the cardinality of the scanned items (step A-4 and A-5).

We further optimize the aforementioned virtual steps (\emph{i.e.}, virtual step A-2.2 and virtual step A-2.3), such that we don't need to really generate any permutation, and thus improve performance. Actually, a permutation is related to a re-arrangement of the elements of an ordered list. If we assign an identifier to each element in the list according to its position, a specific permutation procedure can be represented by an ordered list of these identifiers (\emph{i.e.,} permutation pattern). These permutation patterns are fixed for whatever we permute. We can take advantage of the permutation-pattern to obtain the position of leftmost ``1'' bit in the corresponding permutation without generating it. Specifically, we assign an identifier to each substring of the original hash value. For each substring, we store the index of its leftmost ``1'' bit into an index set (step A-2.2). We can get the index of the leftmost ``1'' bit in any permutation by iterating the index set according to the order of the identifiers in the corresponding permutation pattern (step A-2.3).

We illustrate our LFM sketch by using an simple example. Assume the output of the hash function is 8 bits, which will be spitted into 4 substrings. Let A, B, C, and D be the identifier of each substring respectively. There are 24 potential permutation patterns, including ABCD, ACBD, etc. We keep all the permutation patterns in a $24 \times 4$ matrix (\emph{i.e.,} a permutation-pattern matrix). We then estimate the cardinality of the incoming items following the steps:

\begin{enumerate}[(1)]
\item As for an incoming item ``Na'', we first hash ``Na'' to generate an 8 bit hash value ``00110110''. We then split this hash value into 4 substrings, ``00'', ``11'', ``01'' and ``10'' respectively. Here, A=``00'', ``B=11'', ``C=01'', ``D=10''. We further obtain the index of the leftmost ``1'' bit in each substring A, B, C and D, \emph{i.e.,} -1, 0, 1, 0 (-1 means there is no ``1'' bit in a substring), generating an index set \{-1, 0, 1, 0\}. 
\item To find the leftmost ``1'' bit in a permutation corresponding to a selected permutation pattern ``ACDB'', we don't need to generate the actual permutation (\emph{i.e.,} ``00011011''). Instead, we iterate the index set according to the order of ``ACDB'' until having found a non-negative index. For this case, we start from -1 (\emph{i.e.,} corresponding to ``A''), and then move to 1 (\emph{i.e.,} corresponding to ``C''). Since we have obtained a non-negative index, the iteration ends. We conclude that the index of the leftmost ``1'' bit in permutation ``00011011'' is 3, because  ``-1'' means there is no ``1'' bits in substring ``A'', and ``1'' means there is ``1'' bit in the second bit of substring ``C''. 
\item We aggregate this result into the corresponding bitmap (setting the forth bit in the bitmap to be 1). 
\item Periodically, we get the index of the leftmost ``0'' bit in each bitmap and return the harmonic mean of these indices as the final result.
\end{enumerate}

A detailed description of the LFM sketch is provided in Algorithm 1. On line 2, a new item is hashed, and the generated hash value is split into M substrings. We find the leftmost ``1'' bit in each substring and store the results into an index set (line 4-5). Line 7 begins a loop to calculate the index of the leftmost ``1'' bit in each permutation. For each permutation, in accordance with the sequence of substring identifiers in the corresponding permutation pattern, we iterate the index set to find the first nonegative value and calculate the index of the leftmost ``1'' bit in this permutation (line 8-17). As long as the leftmost ``1'' bit in this permutation is found, we aggregate this result into the corresponding bitmap (line 18). To obtain the estimation result, we search the leftmost ``0'' bit in all the bitmaps (line 21), and calculate the harmonic mean of these positions, obtaining the cardinality estimation of the scanned items (line 22).

\begin{algorithm}
\setlength{\abovedisplayskip}{3pt}
\setlength{\belowdisplayskip}{3pt}
\label{algorithm1}
\algsetup{linenosize=\small}
%\scriptsize
%\fontsize{7pt}{7pt}
%\selectfont
\caption{LFM Sketch}
\label{alg:algorithm1}
\begin{algorithmic}[1]
\renewcommand{\algorithmicrequire}{Input:}
\REQUIRE\
\\\small T: a new item;\\
\small PM[][]: $N \times M$ permutation pattern matrix;
\renewcommand{\algorithmicensure}{Output:}
\ENSURE\
\\\small Est: the estimation of cardinality of the items;\\

\WHILE {The estimation period is not end}
\STATE Hash over T to get a hash value h of size $L$;
\STATE Split h into a set of $M$ substrings;
\STATE Search the index of the leftmost 1 bit in each substring;
\STATE Store the search results into an intex set P[] ; 
\STATE Construct $N$ bitmaps;
\FOR{i = 0 to $N-1$}
\STATE offset = 0;
  \FOR {j = 0 to $M-1$}
     \STATE id = PM[i][j];
     \IF{P[id] != -1}
     \STATE offset = offset + $L$/$M$;
     \ELSE
     \STATE k = offset + P[id];
     \STATE Break;
     \ENDIF
  \ENDFOR
\STATE Set the k-th bit in the i-th bitmap to be 1;
\ENDFOR
\ENDWHILE
\STATE Search the index of the leftmost 0 bit in each bitmap;
\STATE Est is the harmonic mean of the search results; 
\STATE \textbf{Return} Est;
\end{algorithmic}
\end{algorithm}

\begin{table}[!hbp]
\vspace{-13pt}
\caption{Performance Comparison}
\vspace{-6pt}
\begin{tabular}{ccc}
\hline\hline
\textbf{steps} & \small\textbf{Hyperloglog} &\small\textbf{Lightweight} \\
\hline\hline
\small 1 & \small $N\times \delta $ & \small $\delta$ \\
\hline
\small 2 & \small $N\times(2-(0.5)^{L-1})$ & \small $M\times( 2-(0.5)^{\frac{L}{M}-1})+N$\\ 
\hline
\small 3 & \small $N$ & \small $N$\\
\hline
\small 4 & \small $N\times log_{2}\varphi L$ & \small $N\times log_{2}\varphi L$\\
\hline
\small 5 & \small $N$ & \small $N$ \\
\hline
\end{tabular}
\end{table}

\para{Theoretical performance comparison} We compare our LFM sketch with the hyperloglog FM sketch. We assume $\delta$ denotes the computational complexity of a hash operation and  $\varphi$ is an estimation parameter of FM sketch. As it is shown in Table I, our LFM sketch is much more efficient than the hyperloglog FM sketch in two aspects: (1) the computational complexity of step 1 is reduced by a factor of $N$ because the number of the hash operations is 1 in our LFM sketch; (2) the computational complexity of step 2 decreases from $N\times(2-(0.5)^{L-1})$ to $M\times(2-(0.5)^{\frac{L}{M}-1})+N$ by optimizing the leftmost ``1'' bit searching process, since $M$ is exponentially small than $N$ (detailed in Appendix). In summary, the performance of our LFM sketch is significantly improved compared to the hyperloglog FM sketch.

\subsubsection{Detection Scheme}
\label{Detection Scheme}

Based on our LFM sketch, we design ELDA, an efficient and lightweight scheme for detecting CPA attacks in NDN. Similar to the strawman solution, ELDA contains two phases, a traffic monitoring phase and an attack identification phase. 
%\begin{figure}[H]
%  \vspace{-10pt}
%  \centering
%    \includegraphics[width=0.51\textwidth]{./figs/f4}
%  \vspace{-10pt}
%  \caption{The process of ELDA}
%\label{The process of ELDA}
%  \vspace{-8pt}
%\end{figure} 
%\begin{figure}[H]
%  \vspace{-10pt}
%  \centering
%    \includegraphics[width=0.5\textwidth]{./figs/f4}
%  \vspace{-10pt}
%  \caption{The process of ELDA}
%\label{The process of ELDA}
%  \vspace{-8pt}
%\end{figure} 

During the traffic monitoring phase, we apply the LFM sketch to estimate the number of distinct content being requested corresponding to each name prefix. During the CPA identification phase, we compute the interest traffic monitoring result (\emph{i.e.}, the harmonic mean of the indices for the leftmost ``0'' bit in the bitmaps) and compare it with a threshold to detect a potential CPA attack. %Since interest traffic has rough time-of-day pattern, long-term trends, and topology differences \cite{papagiannaki2003long}, we should use an adaptive rather than a fixed threshold. 

In ELDA, we apply a Monte Carlo hypothesis test to estimate a proper CPA detection threshold.
According to Chebyshev's inequality \cite{papoulis2002probability}, the traffic monitoring results will fluctuate around their expectation under an upper bound with high probability. Since this upper bound is related to the expectation and the standard deviation of the recent traffic monitoring results, we can estimate it according to the recent traffic monitoring results so as to get a threshold for distinguishing CPA attack traffic from regular traffic. 
Specifically, we do a Monte Carlo hypothesis test \cite{metropolis1949monte} on the recent traffic monitoring results to verify the detection accuracy of a candidate detection threshold, in which we take a pre-set level of significance to indicate whether a candidate threshold is an accurate estimation of the upper bound or not. If a test result cannot reach the pre-set level of significance, we further enlarge the candidate threshold until we can achieve a qualified test result. We then return the corresponding candidate threshold. Since the Monte Carlo hypothesis test can be performed on a small sample, the size of the sample applied to estimate our detection threshold is reduced and the performance of ELDA is further improved.

\section{Security Analysis}
\label{sec:analysis}

The security of ELDA is captured by the following lemmas, theorems and analysis.

\begin{HNRALemma}
The randomness of the permutations used in LFM sketch is no worse than the hash values generated in FM sketch.
\end{HNRALemma}

\begin{IEEEproof}
(Sketch) FM sketch relies on a hypothesis that each hash value is uniformly distributed over the scalar range or equivalently over the set of binary strings, \emph{i.e.}, every bit in this hash value will be 0 or 1 with the same probability \cite{flajolet1985probabilistic}. When using a hash function to generate a hash value (\emph{i.e.,} a binary string), an intermediate binary string is first partitioned into a number of substrings (\emph{i.e.}, bundles). Certain transformations are then performed on these bundles to propagate randomness further in order to generate a more random binary string. 

The randomness of the binary string can be measured by the branch number of the aforementioned transformation \cite{daemen2002design}. Let $\alpha$ be a vector of bundles, and $\lambda$ be a transformation, then $\lambda(\alpha)=M\alpha$, where M is a matrix used for the transformation. The branch number of $\lambda$ can be computed by $B(\lambda)=\underset{\alpha\neq 0}{min}\{ w(\alpha)+w( M\alpha)\}$, where $w$ is a function used to calculate the bundle weight (\emph{i.e.,} the number of non-zero bundles). In LFM sketch, we partition the hash value into a number of substrings, and generate a large number of permutations of these substrings. Such a permutation procedure is actually a linear invertible transformation (\emph{i.e.,} transformation $\lambda$) on the vector of substrings (\emph{i.e.,} $\alpha$). Thus, for the permutation procedure, we have: $\lambda (\alpha)=M\alpha \neq 0$. Therefore, the permutation procedure (i.e., $\lambda$) will not reduce the branch number of the original hash function, and the randomness of the permutations used in LFM sketch is no worse than the original hash values generated in FM sketch.
\end{IEEEproof}

\begin{HNRATheo}
LFM sketch can achieve approximately the same estimation accuracy as FM sketch.
\end{HNRATheo}

\begin{IEEEproof}
(Sketch) According to \emph{Lemma 1}, the randomness of a permutation used in LFM sketch is no worse than a hash value generated in FM sketch. Although the generated permutations may  contain repeated leftmost substrings, the probability that two permutations contain the same leftmost ``1'' bit is quite small. This probability can be computed as following:
\[Pr = \sum\limits_{k = 1}^M {{{({{({{(\frac{1}{2})}^{\frac{L}{M}}})}^{K - 1}} \cdot (1 - {{(\frac{1}{2})}^{\frac{L}{M}}}))}^2}}  \cdot \frac{1}{{{M^2}}}\]
\[ = \frac{{(1 - {{(\frac{1}{2})}^{\frac{{2L}}{M} \cdot M}})}}{{1 - {{(\frac{1}{2})}^{\frac{{2L}}{M}}}}} \cdot {(1 - {(\frac{1}{2})^{\frac{L}{M}}})^2} \cdot \frac{1}{{{M^2}}} < \frac{1}{{{M^2}}}\] 
where M is the number of the substrings used to form a permutation. 
%$\left(\frac{1}{P_{2}^{M}}\right)^2$
%$Exp <\left(\frac{1}{\binom{M}{2-(1-(0.5)^{\frac{L}{M}})^{M-1}}}\right)^2$, 
According to Theorems in \cite{flajolet1985probabilistic}, 
the relations of the estimation accuracy and the number of the hash values being used in the FM sketch can be formulated as following: $Bias=1+0.31/N$, $SE=0.78/\sqrt{N}$, 
where $N$ is the total number of the hash values generated over a new item, $Bias$ and $SE$ represent the Bias and Standard Error in the estimate process respectively.
To get enough substrings for generating $N$ permutations (N should be larger than 1024~\cite{flajolet1985probabilistic}), M is always larger than or equal to 7. Therefore, the whole randomness reduces little when we use a group of generated permutations instead of a group of original hash values in the estimate process, and thus our LFM sketch can achieve similar estimate accuracy as FM sketch.
\end{IEEEproof}
\para{ELDA can effectively detect CPA attacks} According to Theorem \emph{1}, LFM sketch can achieve nearly the same estimation accuracy as FM sketch. ELDA relies on LFM to monitor the interests possessing a common name prefix, and thus is able to detect the burst of the distinct content requested by these interests accurately. In addition, ELDA can identify a potential CPA attack by using a proper threshold generated by Monte Carlo test.

\section{Evaluations}
\label{sec:evaluations}
In this section, we study the destructive effects of both the LDA attack and the FLA attack. In addition, we evaluate ELDA by comparing it with the most effective approach in the literature.

\vspace{-5pt}
\subsection{Methodology}
\label{methodology}

\begin{figure}[H]
\setlength{\abovedisplayskip}{3pt}
\setlength{\belowdisplayskip}{3pt}
  \vspace{-10pt}
  \centering
    \includegraphics[width=0.45\textwidth]{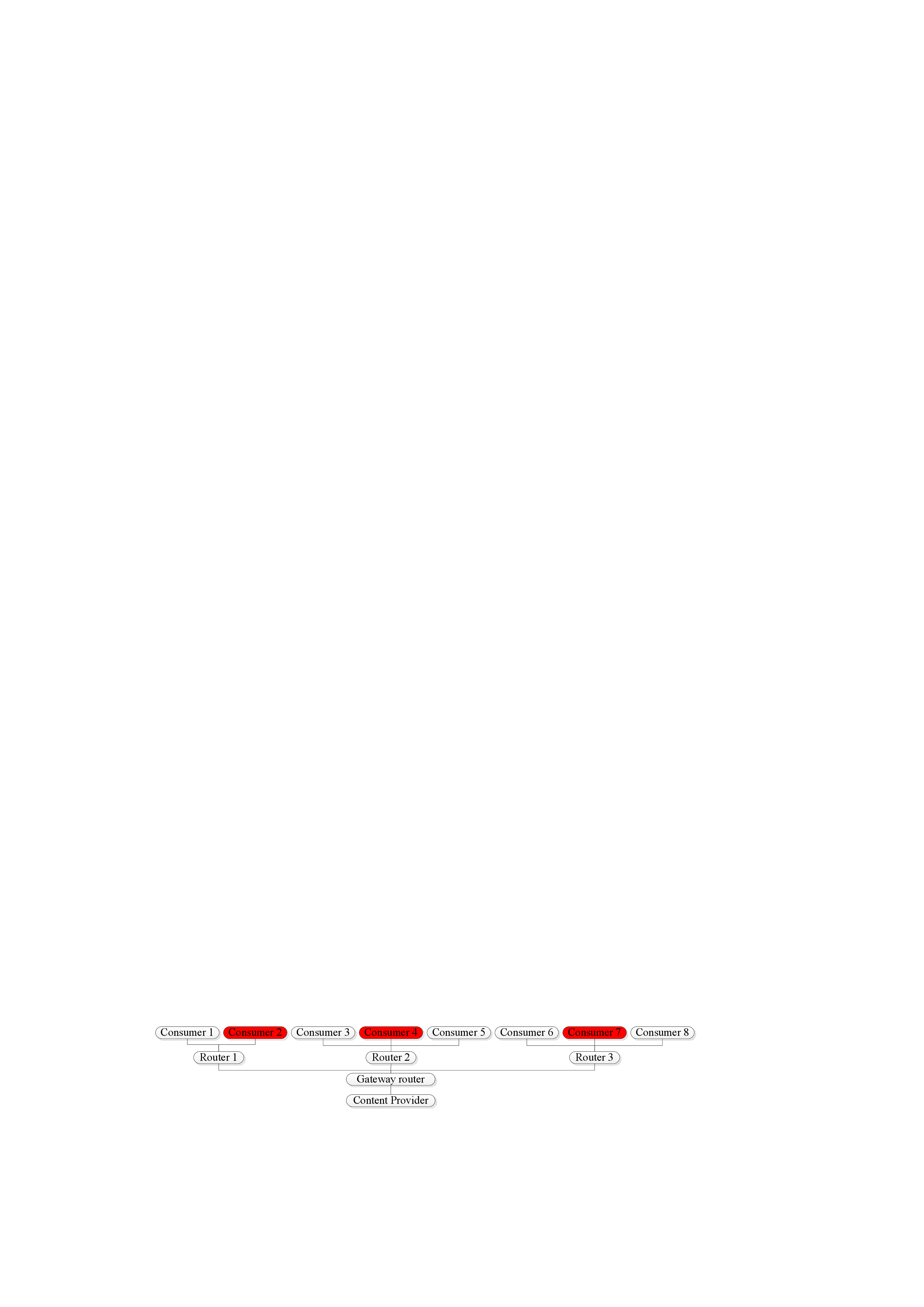}
  %\vspace{-20pt}
  \caption{The network topology}
\label{topology}
  \vspace{-13pt}
\end{figure}

We evaluated the performance of ELDA in ndnSIM \cite{afanasyev2012ndnsim}, a popular NDN simulator. All of our simulations were performed on a local machine, equipped with an Intel Core i7 3.4G CPU and 16G RAM, running Ubuntu 13.04 with kernel version 3.8.8. All the primary parameters we used in ndnSIM are provided in Table II.%\ref{tab:simulation_parameters}.

In our simulations, we deployed a tree-like network topology (see Fig. \ref{topology}). As the common part of real Internet topologies, the tree-liked topology is commonly present in the local networks. CPA attacks traffic will converge at the routers along the critical transmission path, \emph{e.g.}, the gateway routers. In this network topology, we deployed eight consumers, three edge routers, one gateway router and one content provider. We assume consumer 2, 4 and 7 are malicious (\emph{i.e.,} they have been compromised by the attacker and only send malicious interset packets). 

To simulate background traffic, each legitimate consumers sends 3000 regular interests every second. The name of an interest consists of a name prefix (\emph{i.e.,} the first component of the name) and a distinct random number (\emph{i.e.,} the second component of the name). We used five optional name prefixes throughout our simulations: ``/google.com'', ``/amazon.com'', ``/youtube.com'', ``/yahoo.com'' and ``/facebook.com''. We manipulated the random number to distinguish different types of interests (\emph{i.e.,} regular interests, interests requesting unpopular content, and interest requesting nonexistent content), such that a different type of interest will use a random number from a different range. The request frequency of different content by legitimate consumers followed three Zipf-like distributions with different parameters. Their $\alpha$ (the popular content will be requested more frequently with a larger $\alpha$) was 0.7, 0.9 and 1.1 respectively, while the number of the different content was 10000.

The PIT size of each router is 15000, and the cache size is 1000 (which is 1\% of the total number of distinct content used in the simulations). Since a CPA attack can significantly affect the in-network caches using all different types of replacement strategies~\cite{xie2012enhancing}, we chose LRU (least recently used), which is commonly used by the in-network caches~\cite{rosensweig2010approximate}.

To evaluate the damage effect of CPA attacks, all the compromised consumers send malicious interests following our attack model described in Section~\ref{sec:threatmodels}. The name in a malicious interest also consists of a name prefix and a random number. In our simulations, the attacker will use ``/yahoo.com'' as the name prefix. 

\begin{table}[!hbp]
\setlength{\abovedisplayskip}{3pt}
\setlength{\belowdisplayskip}{3pt}
\vspace{-13pt}
\label{tab:simulation_parameters}
\caption{Simulation Parameters}
\vspace{-8pt}
\begin{tabular}{ll}
\hline\hline
\textbf{Simulation Parameters} & \textbf{Value} \\
\hline\hline
\small\# Content items & \small 10,000 \\
\hline
\small Link (Gateway to Content provider) & \small 500Mb/s\\
\hline
\small Link delay (Gateway to Content provider) & \small 20ms\\
\hline
\small Other link & \small 50Mb/s\\
\hline
\small Other link delay & \small 3-5ms\\
\hline
\small CS strategy & \small LRU \\
\hline
\small CS size & \small 1000 content items \\
\hline
\small Maximum PIT size & \small 15000 entries \\
\hline
\small Size of each content item & \small 1,024 bytes \\
\hline
\small Rate of legitimate users & \small 3000 Interests/s \\
\hline
\small Duration for Interests & \small 0-60s \\
\hline
\hline
\end{tabular}
\vspace{-10pt}
\end{table}

\begin{table}[!hbp]
\setlength{\abovedisplayskip}{3pt}
\setlength{\belowdisplayskip}{3pt}
\label{tab:simulation_parameters}
\centering
\caption{Simulation Parameters}
\vspace{-8pt}
\begin{tabular}{lll}
\hline\hline
\textbf{Attack} & \textbf{Zipf} & \textbf{Attack Intensity} \\
\hline\hline
\small\ LDA1,FLA1 & \small 0.7 & \small 3,000 Interests/s \\
\hline
\small\ LDA2,FLA2 & \small 0.7 & \small 6,000 Interests/s \\
\hline
\small\ LDA3,FLA3 & \small 0.9 & \small 3,000 Interests/s \\
\hline
\small\ LDA4,FLA4 & \small 0.9 & \small 6,000 Interests/s \\
\hline
\small\ LDA5,FLA5 & \small 1.1 & \small 3,000 Interests/s\\
\hline
\small\ LDA6,FLA6 & \small 1.1 & \small 6,000 Interests/s\\
\hline
\hline
\end{tabular}
\vspace{-10pt}
\end{table}

\para{Implementation of ELDA} When implementing ELDA, we used murmur hash function \cite{appleby2008murmurhash} to generate hash values due to its efficiency. Each interest name is hashed to a 32-bit hash string, which is further split into 8 substrings. In this way, we are able to estimate the cardinality of the requested content accurately, since we are allowed to use at most 8! bitmaps. In our implementation, we randomly selected 256 bitmaps. When using Monte Carlo hypothesis test to obtain the threshold, our significance level is 0.5\%, and the sample size is 10.

\para{Implementation of a baseline scheme} To show the effectiveness of ELDA, we also implemented the most efficient detection scheme \cite{conti2013lightweight} in the literature as a baseline for comparison. This scheme monitors the interest traffic, and calculates the number of each newly requested content. It identifies a CPA attack by checking whether the requesting frequency of a recently requested content deviates from a threshold, which was obtained based on the expectation and the variance of the past requesting frequency for this content.

\subsection{Damage Effect of LDA and FLA Attacks}
\label{damage effect}

We evaluated the damage effect of both the LDA and the FLA attacks in order to justify our threat model.

\begin{figure*}
\centering
\begin{tabular}[t]{lcccr}
\subfigure[LDA:Cache hit rate]{{\includegraphics[width=0.18\textwidth]{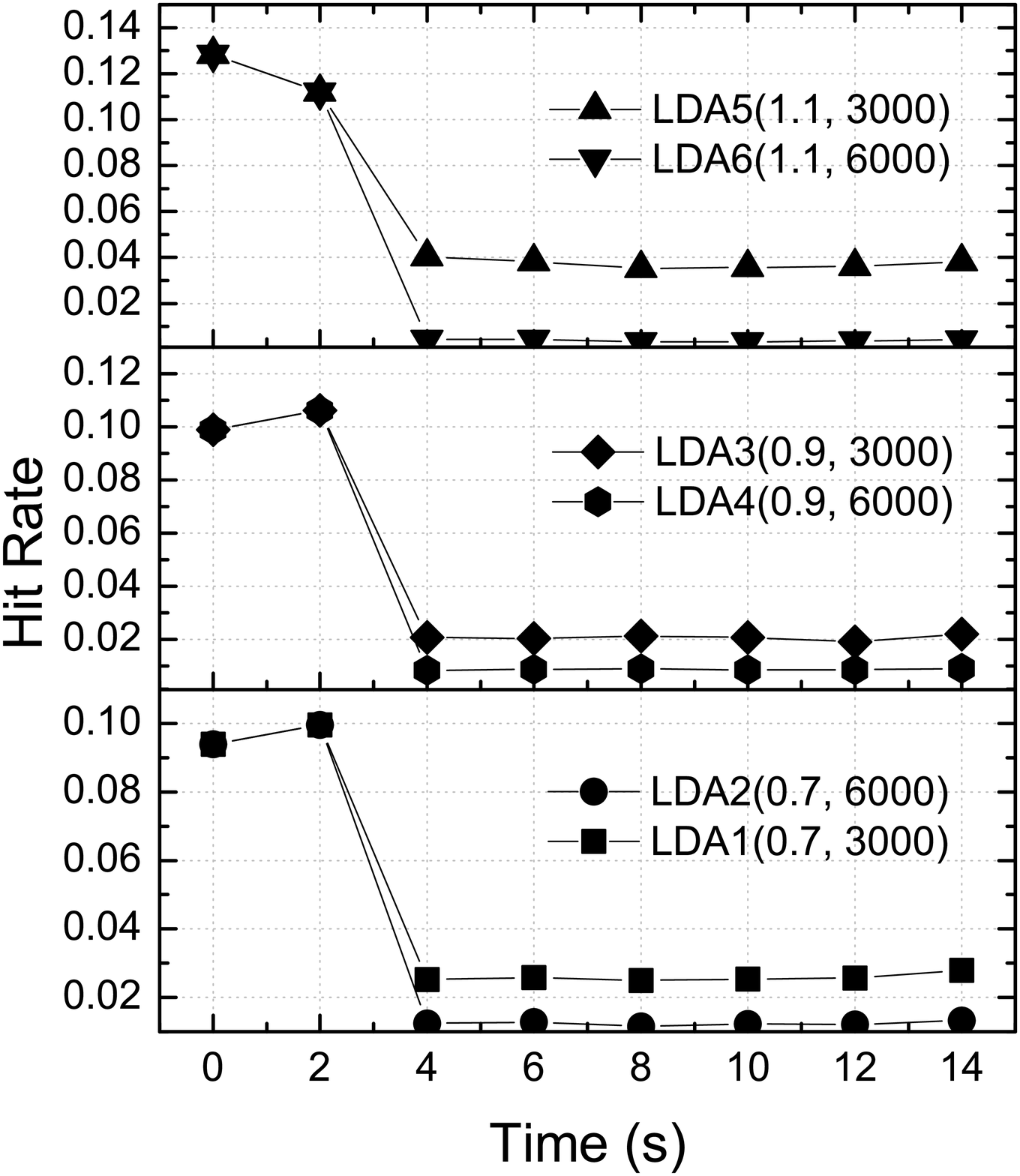}}
\label{fig:LDA:Cache Hit Rate}}
%\vspace{-13pt}
\subfigure[LDA: Average RTT ]{{\includegraphics[width=0.18\textwidth]{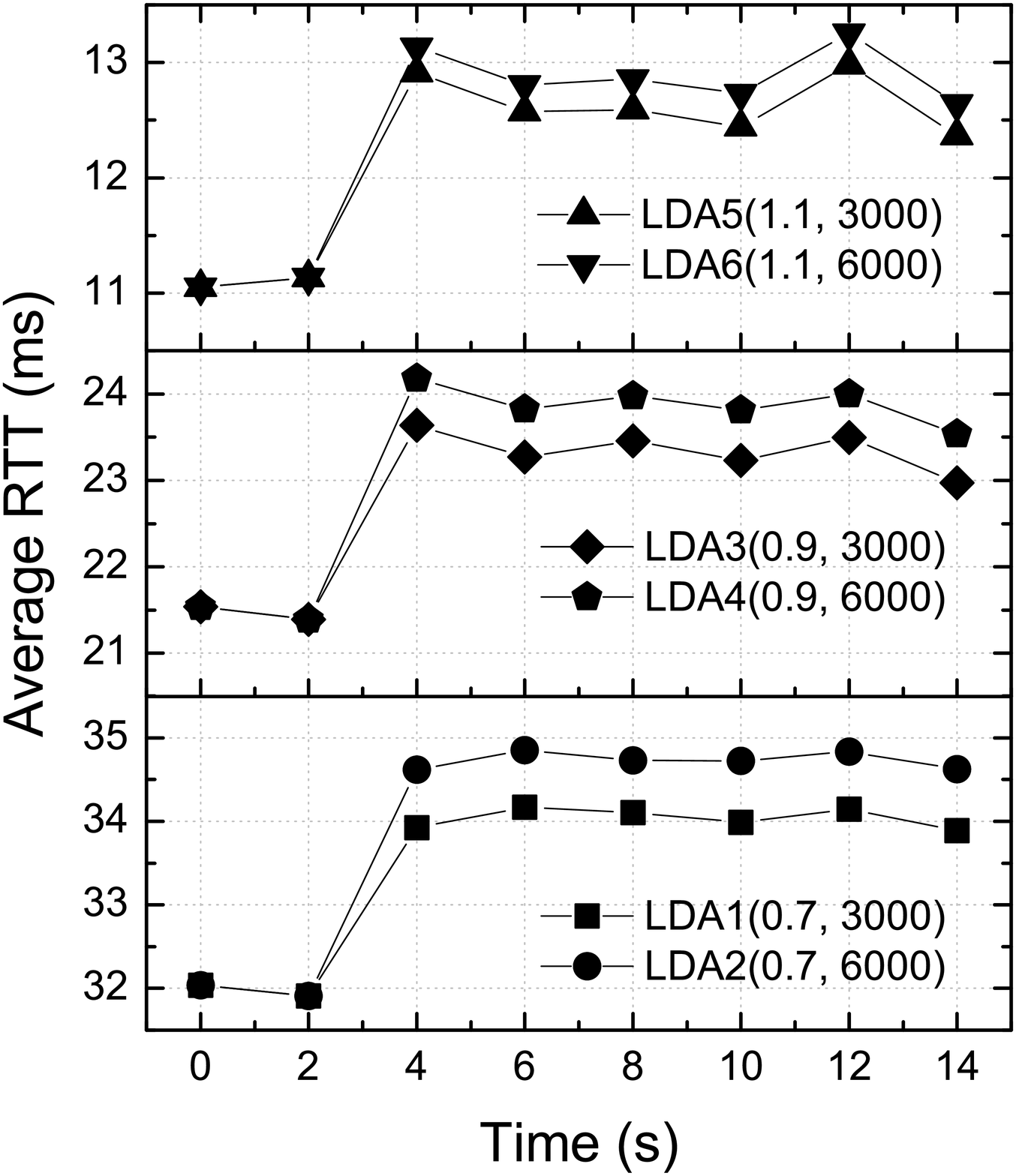}}
\label{fig:LDA:Average RTT}}
%\vspace{-13pt}
\subfigure[FLA: Cache hit rate]{{\includegraphics[width=0.18\textwidth]{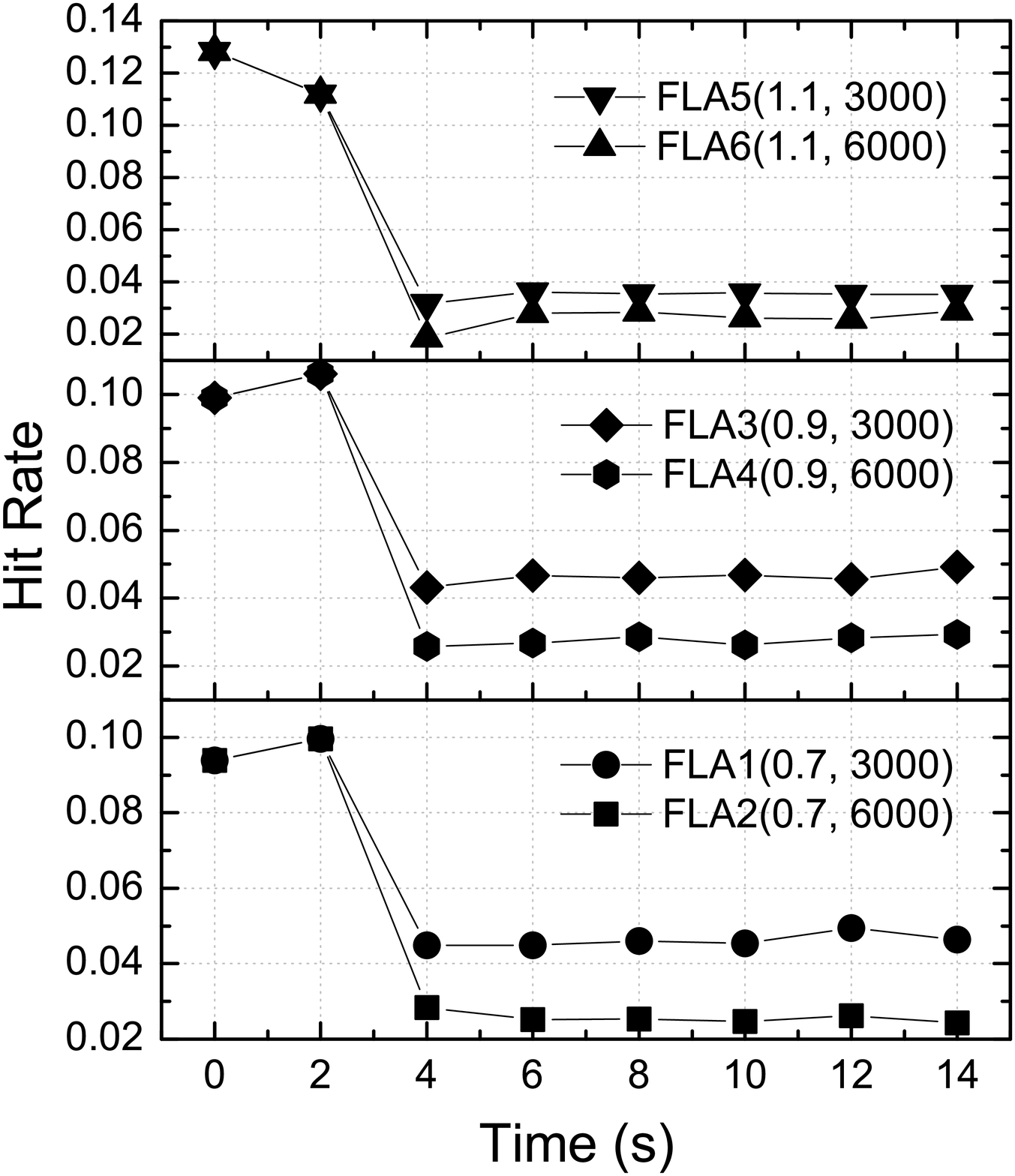}}
\label{fig:FLA:Cache Hit Rate}}
\subfigure[FLA: PIT avaliable rate]{{\includegraphics[width=0.18\textwidth]{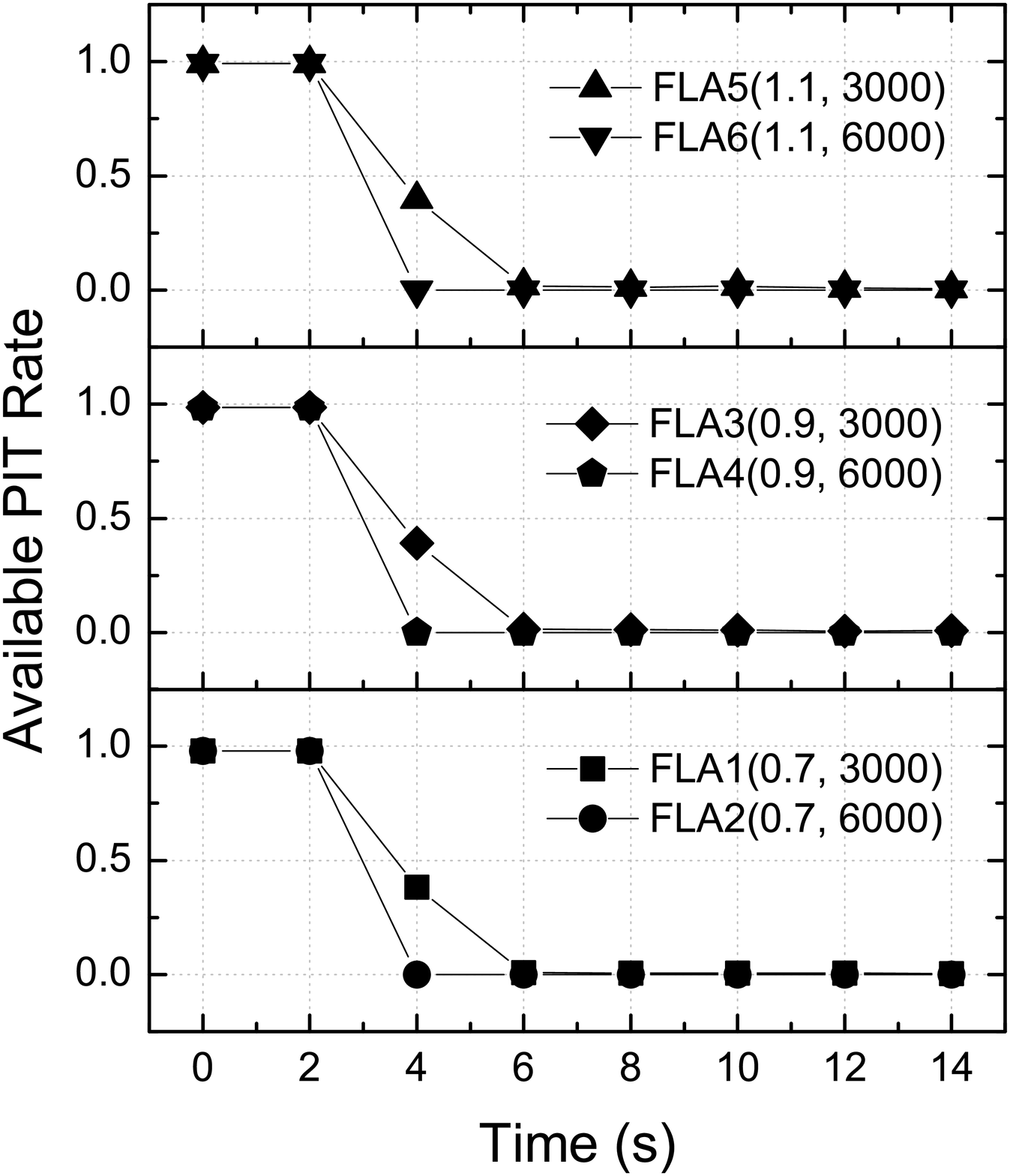}}
\label{fig:FLA:PIT Avaliable Rate}}
\subfigure[FLA: Average RTT]{{\includegraphics[width=0.18\textwidth]{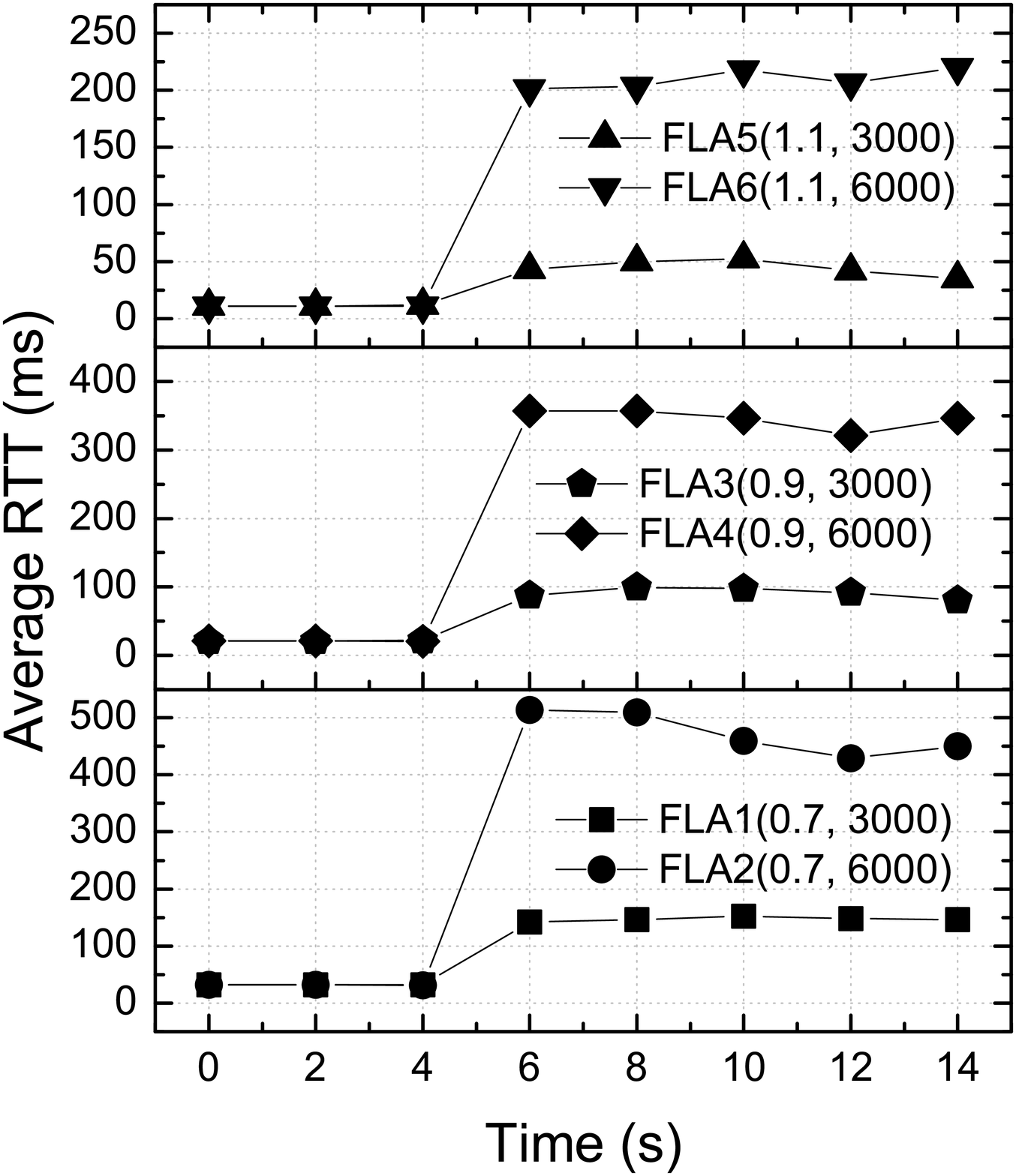}}
\label{fig:FLA:Average RTT}}
\end{tabular}
%\hspace{1cm}
%\subfigure[RA-BF versus RA-PAC with different scale]{\includegraphics[scale=0.4]{./figs/f17}
%\label{fig:RA-BF_vs_RA-PAC}}
\vspace{-8pt}
\caption{The effect of the CPA attacks}
\label{fig:LDAs & FLAs}
\end{figure*}

\para{LDA attacks} We simulated six LDA attacks, LDA1, LDA2, LDA3, LDA4, LDA5 and LDA6. As listed in Table III, each compromised consumer sends malicious interests per second to request unpopular content, while the regular traffic follows a Zipf-like distribution with a specific $\alpha$. In these attacks, the compromised consumers send out malicious interests from second 2. We have two observations: (1) In LDA2, LDA4 and LDA6, the CS hit rate in the gateway router drops almost to 0, and that in LDA1, LDA3 and LDA5 also declines significantly (see Fig. \ref{fig:LDA:Cache Hit Rate}). In LDA5, the number of the attack requests is only equal to the number of the regular requests and the larger $\alpha$ (\emph{i.e.,} 1.1) is used to generate regular traffic in this case. As a result, the cache hit rate maintains at a higher level compared to the simulation results of the other LDA attacks. (2) When an LDA attack happens, the average round-trip time (RTT) of regular interests increases 2 to 3 milliseconds in terms of different $\alpha$ of the regular traffic and attack frequency of LDA attacks. When we used a large $\alpha$ and low malicious request frequency in the simulation, the average RTT increases more slowly (see Figure \ref{fig:LDA:Average RTT}). We conclude that, the LDA attacks affect the overall performance of NDN network more seriously while sending more malicious interests per second. On the other hand, the LDA attacks degrade the performance less when the $\alpha$ of the regular traffic is large.

\para{FLA attacks} We also simulated the damage effect of six FLA attacks, FLA1, FLA2, FLA3, FLA4, FLA5 and FLA6. The compromised consumers request unpopular content from the second 2 and request non-existent content from the second 3. The name prefixes of these malicious interests all are ``/yahoo.com’’. We have three observations: Firstly, when the compromised consumers send more malicious interests per second, the performance of the gateway router is degraded significantly. Secondly, the FLA attacks have the damage effect on the PIT (see Fig.\ref{fig:FLA:PIT Avaliable Rate}), \emph{i.e.,} the available space of PIT declines to 0 and no more subsequent interests can be forwarded. In this way, the attacker keeps the unpopular content in the CS by blocking the subsequent content, and thus the CS hit rate significantly degrades. Thirdly, the FLA attacks affect the gateway router more significantly compared to the LDA attacks (see Fig.\ref{fig:FLA:Average RTT}). The FLA attacks delay the average round-trip time much more compared to the LDA attacks (see Fig.\ref{fig:LDA:Average RTT}).

\subsection{Evaluations of ELDA}
\label{Evaluations of ELDA}

In this section, we compare ELDA with the baseline scheme in terms of effectiveness, efficiency and resource consumption. We will show that compared to the baseline scheme, ELDA maintains the same effectiveness while having much better efficiency and significantly less resource consumption.

\begin{figure}[H]
\setlength{\abovedisplayskip}{3pt}
\setlength{\belowdisplayskip}{3pt}
\vspace{-13pt}
\centering
\subfigure[Detection rate]{
\label{fig:attack_detect.sub.1}
\raisebox{-10pt}{\includegraphics[width=0.23\textwidth]{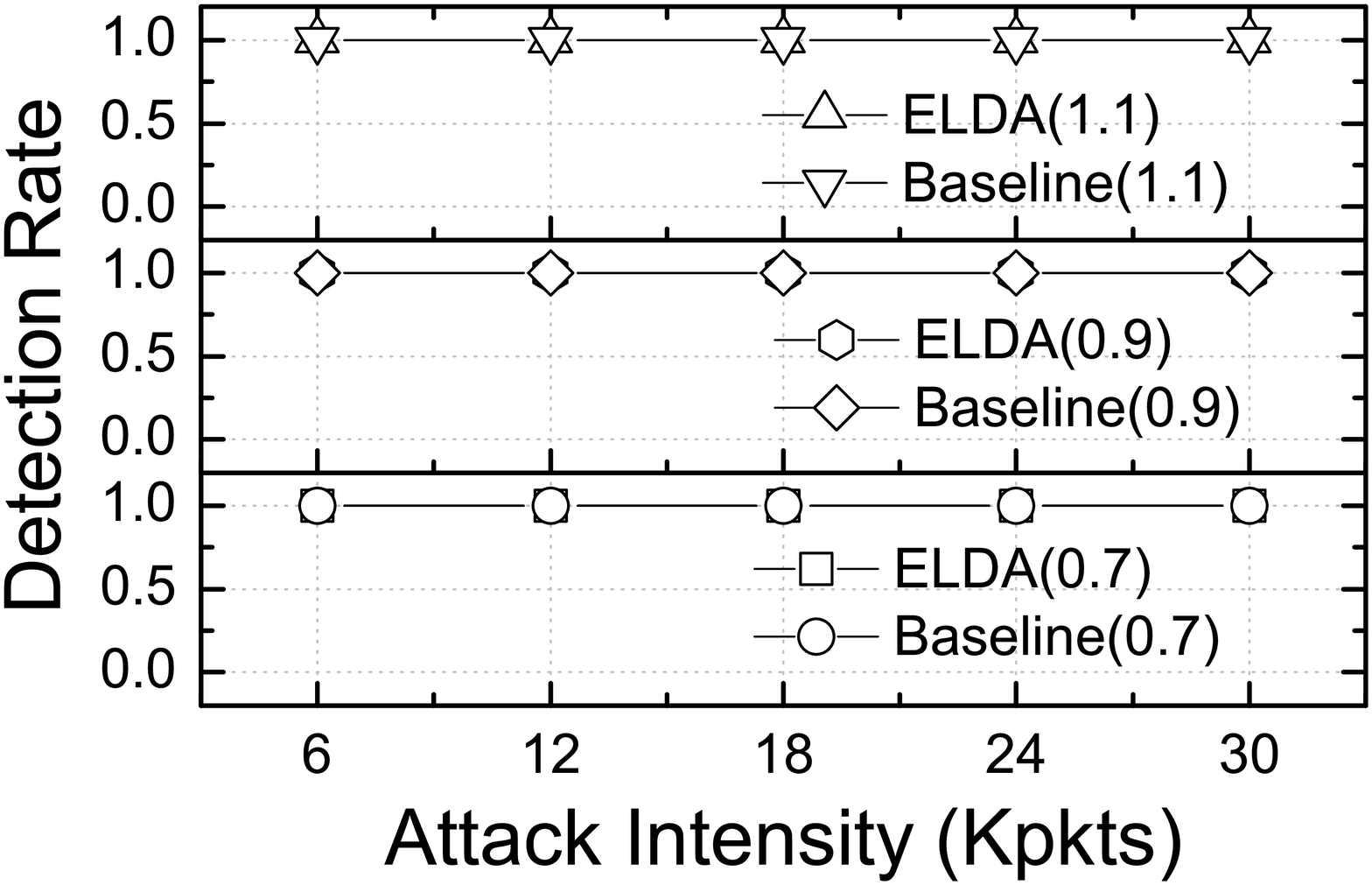}}}
\subfigure[False positive rate]{
\label{fig:attack_detect.sub.2}
\raisebox{-10pt}{\includegraphics[width=0.23\textwidth]{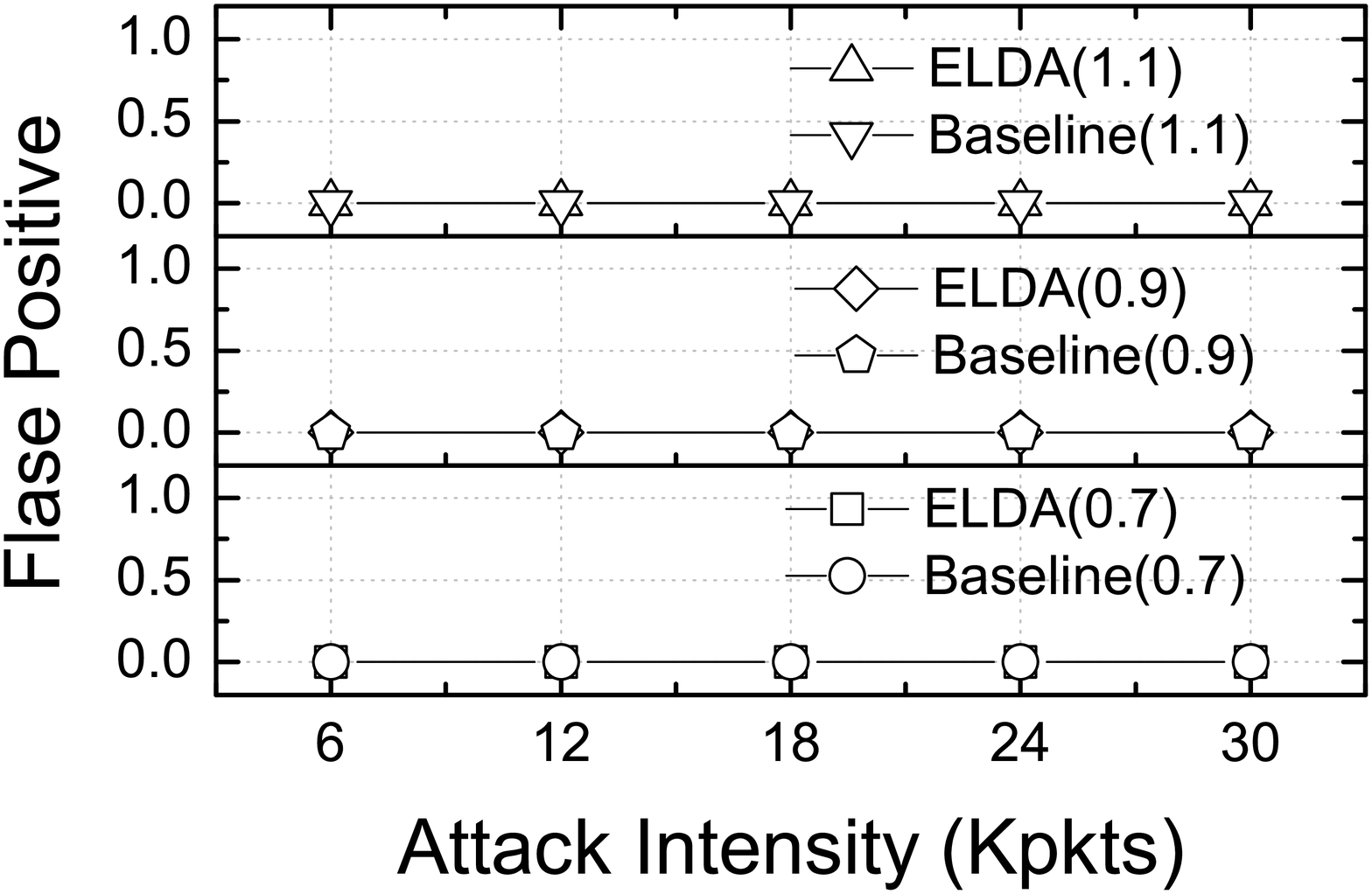}}}
\vspace{-8pt}
\caption{Attack Detection}
\label{fig:attack_detect}
\vspace{-15pt}
\end{figure}

\para{The effectiveness of ELDA} To show the effectiveness of ELDA, we evaluate both the detection rate and the false positive rate, while performing a series of damage FLA attacks (the attack intensity is measured in terms of kilo packets per second, \emph{i.e.,} Kpkts). The detection rate is defined as a ratio between the number of correctly detected CPA attacks and the actual number of CPA attacks. A false positive may happen if the detection scheme falsely classifies a non-attack event as an attack event, and we define the false positive rate as a ratio between the number of false positives and the total number of CPA detection events. In Fig. \ref{fig:attack_detect}, we have two observations: (1) the detection rates of both ELDA and the baseline scheme are 100\%, regardless of $\alpha$ and the malicious request frequency; (2) the false positive rates of both schemes are zero. We conclude that both schemes can detect CPA attacks effectively.

\begin{figure}[H]
\setlength{\abovedisplayskip}{3pt}
\setlength{\belowdisplayskip}{3pt}
\vspace{-8pt}
\centering
\subfigure[Processing time]{
\label{fig:com.sub.1}
 \raisebox{-12pt}{\includegraphics[width=0.235\textwidth]{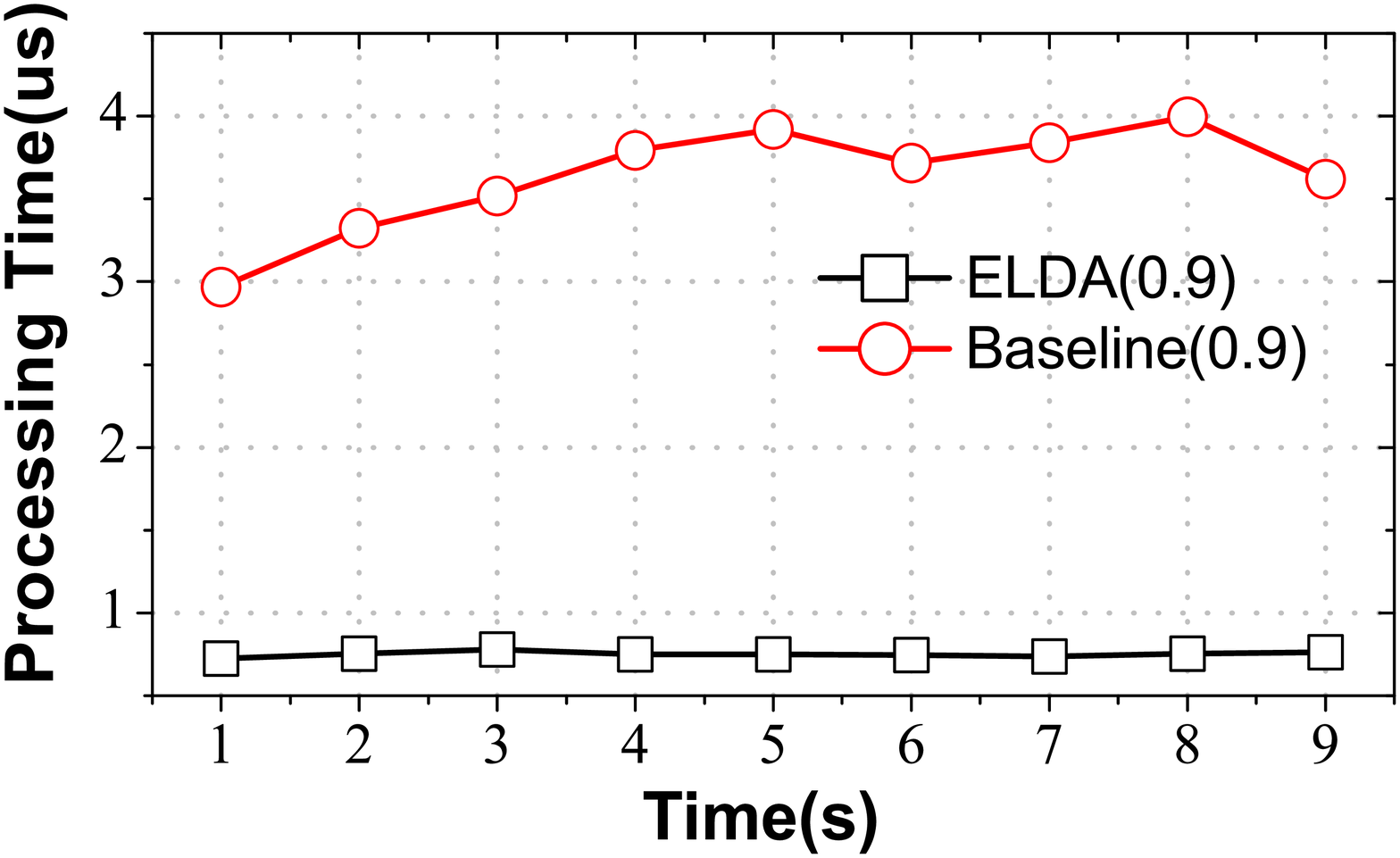}}}
\subfigure[Throughput]{
\label{fig:com.sub.2}
 \raisebox{-12pt}{\includegraphics[width=0.23\textwidth]{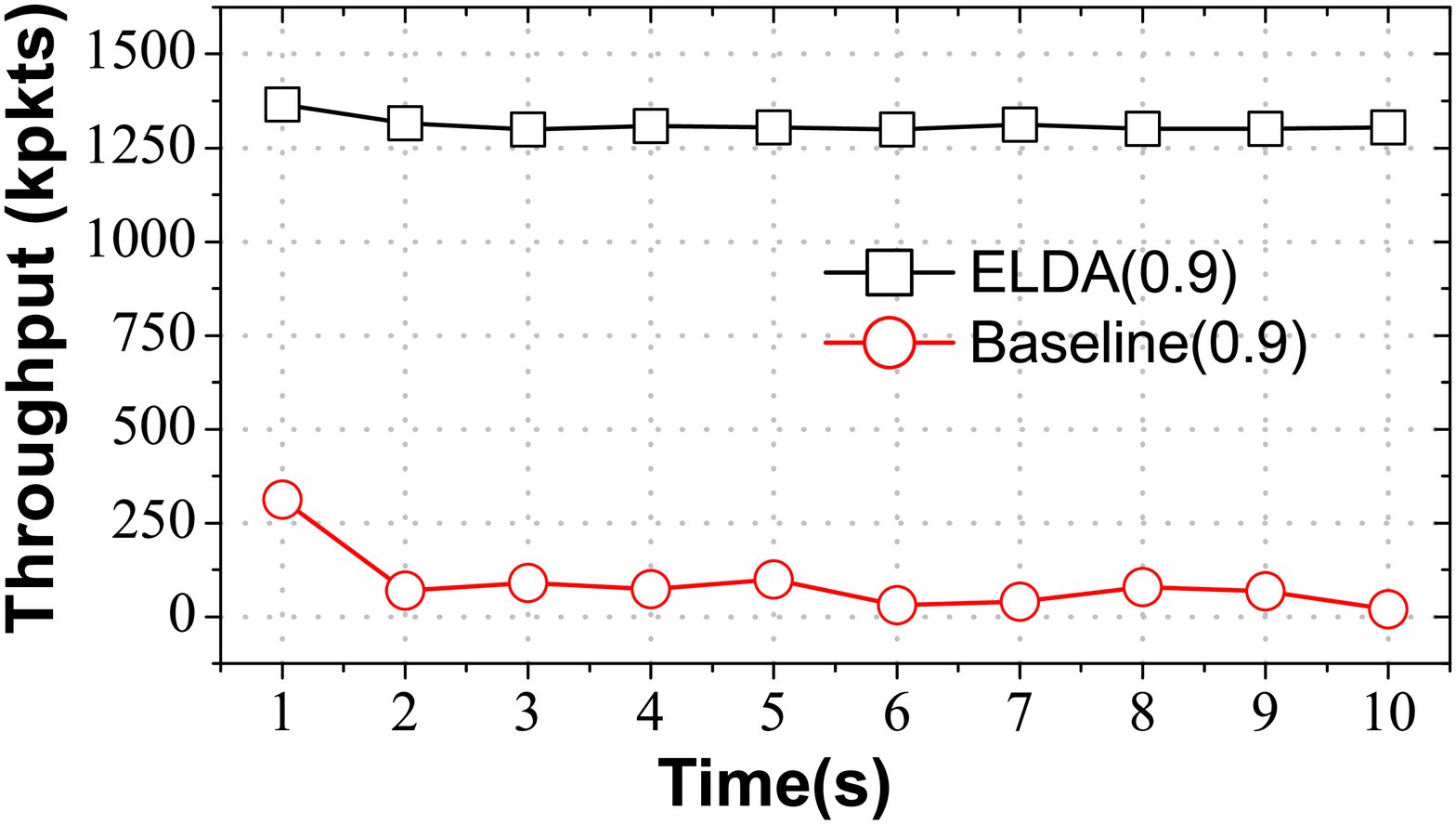}}}
\vspace{-10pt}
\caption{Efficiency}
\label{fig:com}
\vspace{-15pt}
\end{figure}

\para{The efficiency of ELDA} We evaluate the efficiency of ELDA in terms of the processing time and the network throughput. This is measured in terms of kilo packets per second, \emph{i.e.,} Kpkts. Since the processing time and the network throughput are both calculated regardless of the type of request packets (requesting popular content or unpopular content, and being a regular request or a malicious request), we did not discriminate the type of requests and thus set the $\alpha$ value to 0.9 here. As shown in Fig.~\ref{fig:com.sub.1}, the baseline scheme needs approximately 4 $\mu$s to handle a newly arriving interest, while ELDA only needs approximately 0.8 $\mu$s. Moreover, the baseline scheme can only process less than 200,000 interest packets per second, while ELDA is able to process 1,300,000 interest packets per second(see Fig.\ref{fig:com.sub.2}). We conclude that ELDA is much more efficient than the baseline scheme.

\begin{figure}[H]
\setlength{\abovedisplayskip}{3pt}
\setlength{\belowdisplayskip}{3pt}
\vspace{-12pt}
\centering
\subfigure[CPU usage rate]{
\label{fig:cpu.sub.1}
\raisebox{-12pt}{\includegraphics[width=0.23\textwidth]{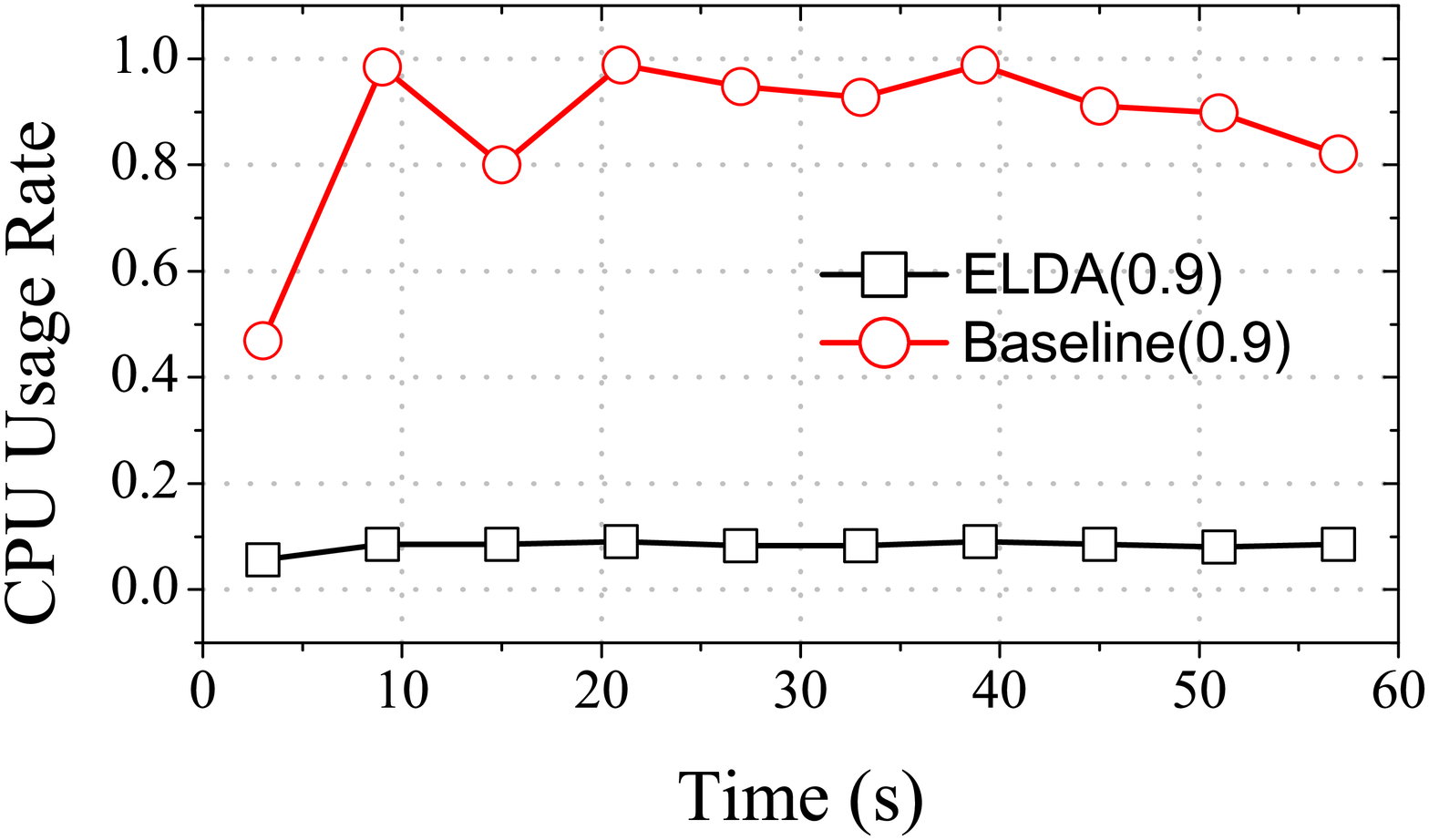}}}
\subfigure[Memory usage]{
\label{fig:cpu.sub.2}
\raisebox{-12pt}{\includegraphics[width=0.23\textwidth]{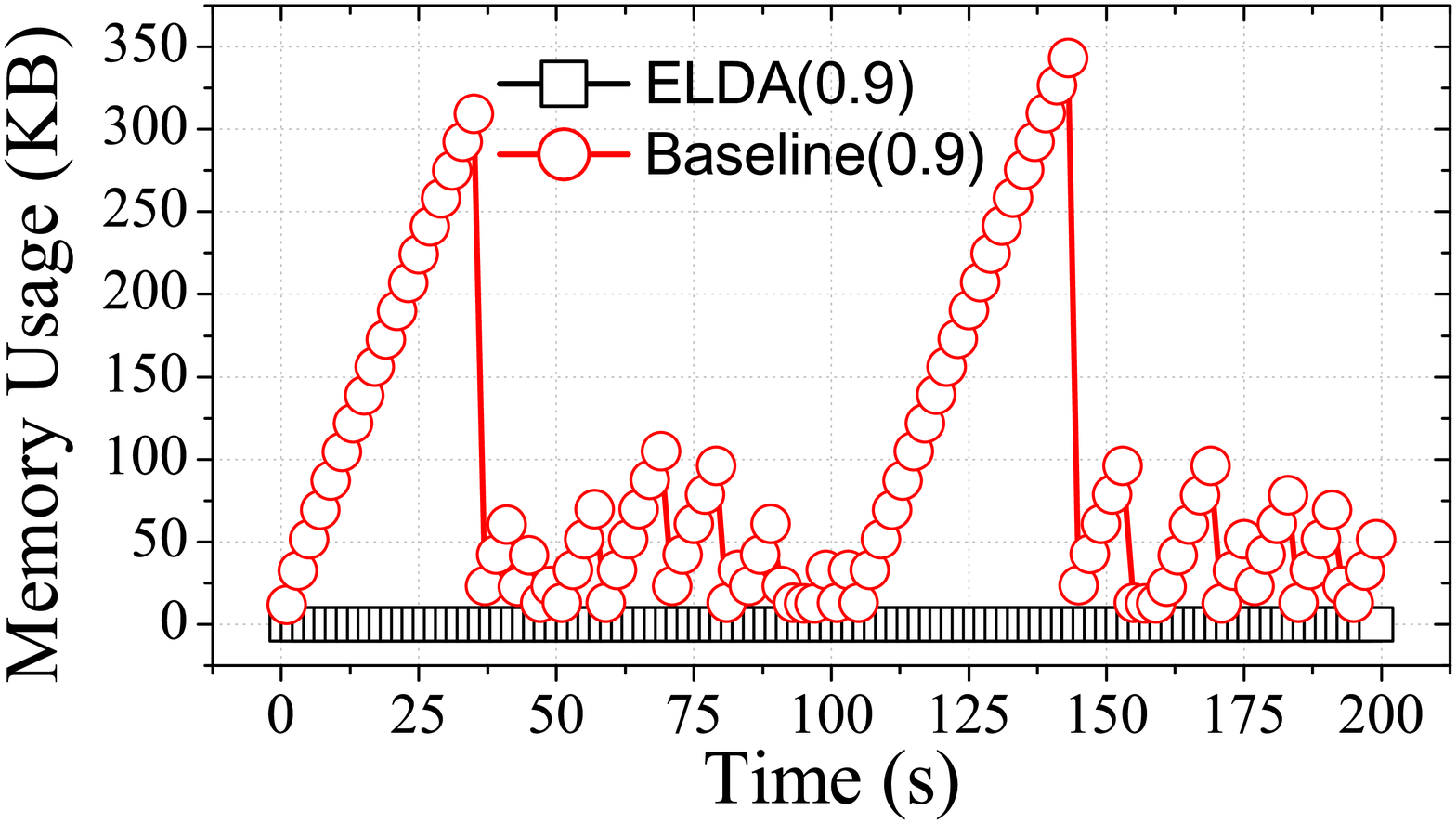}}}
\vspace{-8pt}
\caption{Resource consumption}
\label{fig:cpu}
\vspace{-10pt}
\end{figure}

\para{Resource consumption of ELDA} To evaluate the resource consumption of ELDA, we traced ELDA's CPU and memory usage (see Fig.\ref{fig:cpu}). Similarly to the efficiency simulation, we set the $\alpha$ value to 0.9. When monitoring interest traffic, the baseline scheme almost monopolizes the CPU, while ELDA only uses 10\% of the CPU. Meanwhile, the baseline scheme requires more than 100KB memory to monitor a set of interests, each of which possesses the same name prefix. In contrast, ELDA only requires 488 bits. To identify the prefix used by attackers and throttle the corresponding malicious interests, a router usually needs to monitor interests with different name prefixes simultaneously. To achieve this, the baseline scheme may require much more amount of memory. For example, if the router needs to simultaneously handle 10,000 different name prefixes, the overall memory required in the baseline scheme will be 1GB, while in ELDA, it will be only 5MB. Compared to the baseline scheme, ELDA requires significantly less computation and memory resources.

\section{Related Work}
\label{sec:related}

The CPA attack is a major threat to in-network caches in the current Internet \cite{gupta1990improving}. Numerous research efforts \cite{zhuang2003hardware, gao2006internet} have been devoted to mitigate this type of attack. In particular, they analyze the characteristics of the network flows in terms of different IP addresses in a certain period, and thus are able to identify and throttle a malicious client. However, in NDN, we need to analyze the network traffic in terms of hierarchically structured packet names, rather than well-formatted IP addresses. Due to the differences between packet names and IP addresses in nature, we need to re-design the CPA detection mechanism for NDN networks \cite{xie2012enhancing}.

Although CPA attacks will degrade the performance of NDN networks significantly \cite{Matthias2013Backscatter,xie2012enhancing}, till now, there is very limited work on mitigating CPA attacks in NDN. Xie et al. \cite{xie2012enhancing} propose a proactive CPA countermeasure to enhance cache robustness to CPA attacks without any attack detection process. In their work, when a router receives a content packet, it evaluates an exponential function of the requesting frequency to provide the probability that this content packet will be kept in the cache. One disadvantage of their work is, it will prevent caching some potential popular content requested by regular users. Conti et al. \cite{conti2013lightweight} propose a lightweight approach for detecting LDA attacks. They first point out that the existing proactive countermeasure \cite{xie2012enhancing} is ineffective when defending against realistic adversaries. They then propose their approach based on the hit rate records of the received content. Their solution will consume a large number of computation and memory resources for storing and iterating these hit rate records. 
%To detect CPA attacks, Park et al. \cite{park2012detection} apply a statistical sequential analysis, which relies on matrix analysis technology to estimate network aberrant behavior. 
%In \cite{karami2015anfis}, Karami et al. propose an ANFIS-based cache replacement method, in which they apply both Artificial Neural Network (ANN) and fuzzy algebra to detect CPA attacks. Both the aforementioned approaches are not efficient enough in a realistic NDN network because they all build on top of heavyweight primitives.
Some other approaches \cite{park2012detection,karami2015anfis} are also not efficient enough in a realistic NDN network because they all build on top of heavyweight primitives like artificial neural network.
\section{Conclusion}
\label{sec:conclusion}
To mitigate the CPA attacks in NDN, we propose ELDA, an efficient and lightweight CPA detection scheme. We first analyze the cache pollution attacks in NDN, discovering a common attack symptom for both of the LDA attack and the FLA attack. In addition, we design a novel lightweight FM sketch (LFM sketch) which can capture the attack symptom in the interest traffic. Finally, we propose ELDA based on the LFM sketch. Both the theoretical analysis and the simulations demonstrate ELDA could effectively and efficiently detect CPA attacks in NDN.

\section{Acknowledgements}
Our work is supported by National Program on Key Basic Research Project (2012CB315804), and by NSFC Grant (60803139, 61173133, 61133015). 
Bo Chen is supported in part by NSF grants (NSF CNS-1161541, NSF CNS-1318572, NSF CNS-1223239, NSF CCF-0937833), by US ARMY award W911NF-13-1-0142, and by gifts from Northrop Grumman Corporation, Parc/Xerox, and Microsoft Research.

%{\scriptsize
%\bibliographystyle{abbrv}
%\small{
%\bibliography{mainref}
%}
\bibliographystyle{unsrt}
\bibliography{mainref2}

\begin{thebibliography}{10}

\bibitem{gantz2012digital}
John Gantz and David Reinsel.
\newblock The digital universe in 2020: Big data, bigger digital shadows, and
  biggest growth in the far east.
\newblock Technical report, EMC, 2012.

\bibitem{roberts2009clean}
James Roberts.
\newblock The clean-slate approach to future internet design: a survey of
  research initiatives.
\newblock {\em Annals of telecommunications}, 64(5):271--276, 2009.

\bibitem{jacobson2009networking}
Van Jacobson, K~Smetters, and et~al.
\newblock Networking named content.
\newblock In {\em CONEXT}. ACM, 2009.

\bibitem{Zhang2010data}
Zhang Lixia, Van Jacobson, and et~al.
\newblock Named data networking (ndn) project.
\newblock Technical report, PARC, 2010.

\bibitem{gupta1990improving}
Rajiv Gupta and Chi-Hung Chi.
\newblock Improving instruction cache behavior by reducing cache pollution.
\newblock In {\em Supercomputing}. IEEE, 1990.

\bibitem{gao2006internet}
Yan Gao, Leiwen Deng, and et~al.
\newblock Internet cache pollution attacks and countermeasures.
\newblock In {\em ICNP}. IEEE, 2006.

\bibitem{xie2012enhancing}
Mengjun Xie, Indra Widjaja, and et~al.
\newblock Enhancing cache robustness for content-centric networking.
\newblock In {\em INFOCOM}. IEEE, 2012.

\bibitem{conti2013lightweight}
Mauro Conti, Paolo Gasti, and et~al.
\newblock A lightweight mechanism for detection of cache pollution attacks in
  named data networking.
\newblock {\em Computer Networks}, 57(16):3178--3191, 2013.

\bibitem{karami2015anfis}
Amin Karami and Manel Guerrero-Zapata.
\newblock An anfis-based cache replacement method for mitigating cache
  pollution attacks in named data networking.
\newblock {\em Computer Networks}, 80:51--65, 2015.

\bibitem{afanasyev2012ndnsim}
Alexander Afanasyev, Ilya Moiseenko, and et~al.
\newblock ndnsim: Ndn simulator for ns-3.
\newblock Technical report, UCLA, 2012.

\bibitem{flajolet1985probabilistic}
Philippe Flajolet and G~Nigel Martin.
\newblock Probabilistic counting algorithms for data base applications.
\newblock {\em Journal of computer and system sciences}, 31(2):182--209, 1985.

\bibitem{flajolet2008hyperloglog}
Philippe Flajolet, {\'E}ric Fusy, and et~al.
\newblock Hyperloglog: the analysis of a near-optimal cardinality estimation
  algorithm.
\newblock {\em DMTCS}, (1), 2008.

\bibitem{metropolis1949monte}
Nicholas Metropolis and Stanislaw Ulam.
\newblock The monte carlo method.
\newblock {\em Journal of the American statistical association},
  44(247):335--341, 1949.

\bibitem{gasti2013and}
Paolo Gasti, Gene Tsudik, and et~al.
\newblock Dos and ddos in named data networking.
\newblock In {\em ICCCN}. IEEE, 2013.

\bibitem{papoulis2002probability}
Athanasios Papoulis and S~Unnikrishna Pillai.
\newblock {\em Probability, random variables, and stochastic processes}.
\newblock Tata McGraw-Hill Education, 2002.

\bibitem{daemen2002design}
Joan Daemen and Vincent Rijmen.
\newblock The design of rijndael. information security and cryptography.
\newblock {\em Text and Monographs, Springer Verlag}, 2002.

\bibitem{rosensweig2010approximate}
Elisha~J Rosensweig, Jim Kurose, and et~al.
\newblock Approximate models for general cache networks.
\newblock In {\em INFOCOM}. IEEE, 2010.

\bibitem{appleby2008murmurhash}
Austin Appleby.
\newblock Murmurhash 2.0.
\newblock http://sites.google.com/site/murmurhash, 2015.

\bibitem{zhuang2003hardware}
Xiaotong Zhuang and H-HS Lee.
\newblock A hardware-based cache pollution filtering mechanism for aggressive
  prefetches.
\newblock In {\em ICPP}. IEEE, 2003.

\bibitem{Matthias2013Backscatter}
Thomas C.~Schmidt Matthias~Wahlisch and et~al.
\newblock Backscatter from the data plane - threats to stability and security
  in information-centric network infrastructure.
\newblock {\em Computer Networks}, 57(16):3192--3206, 2013.

\bibitem{park2012detection}
Hyundo Park, Indra Widjaja, and et~al.
\newblock Detection of cache pollution attacks using randomness checks.
\newblock In {\em ICC}. IEEE, 2012.

\end{thebibliography}

\appendix
\section{Appendix}
\label{sec:appendix}
\subsection{Analysis of the Computational Complexity of the Leftmost ``1'' Bit Search} 
\label{sec: computational complexity of search }
We analyze the computational complexity of the leftmost ``1'' bit search of FM sketch and LFM sketch as following. To find the leftmost ``1'' bit in a hash value, FM sketch searches the hash value from the leftmost bit, and the computational complexity is 
$\sum\limits_{k = 1}^L {k \cdot {( {\frac{1}{2}} )}^K}$, 
%\[2 \cdot C(L) = \sum\limits_{k = 1}^L {k \cdot {( {\frac{1}{2}} )}^{K - 1}}, \]
where L is the length of the hash value. We define a function for formulating this expression: 
\[L(x) = \sum\limits_{k = 1}^L {k \cdot {x}^{K - 1}}. \]
Then we can compute it as following:
\[L(x) = {( \int {L(x)} dx )^\prime } = \frac{{x - {x^L}}}{{{(1 - x)}^2}}.\]
%Plugging $\frac{1}{2}$ into $L(x)$, we obtain,
%\[C(L) = 2 - {( {\frac{1}{2}} )^{L - 1}}.\]
Therefore, considering the number of the hash values (\emph{i.e.,} N), the total computational complexity of the leftmost ``1'' bit search in FM sketch is 
$N \cdot ( {2 - {{( {\frac{1}{2}} )}^{L - 1}}} )$.

In LFM, the computational complexity of finding the leftmost ``1'' bit in a generated permutation consists of two parts. The first part is the computational complexity of finding the leftmost ``1'' bit in a hash substring. Similarly, it is equal to 
$2 - {( {\frac{1}{2}} )^{\frac{L}{M} - 1}}$, 
where M is the number of the substrings.
The second part is the computational complexity of finding the leftmost ``1'' bit in a series of hash substrings that compose a permutation used in LFM sketch. In this process, we find the first nonzero hash substring and return its leftmost ``1'' bit as a result. In a similar way, when $\frac{L}{M}>2$, the computational complexity of finding such a substring is equal to   
$\sum\limits_{k = 1}^M {k{{({{(\frac{1}{2})}^{\frac{L}{M}}})}^{K - 1}} \cdot (1 - {{(\frac{1}{2})}^{\frac{L}{M}}})}$, which is less than 1.
Therefore, in summary, the  computational complexity of finding the leftmost ``1'' bit in a generated permutation is less than 
$M \cdot ( {2 - {{( {\frac{1}{2}} )}^{\frac{L}{M} - 1}}} ) + N$.

%\begin{spacing}{1.1}
%{\scriptsize
%\bibliographystyle{unsrt}
%\bibliography{mainref}
%}
%\end{spacing}

%}

\clearpage
\end{document}